\documentclass[useAMS,usenatbib,referee]{biom}
\usepackage{graphicx}

\let\theoremstyle\relax

\usepackage{amsmath,amsfonts,amsthm,bm} 
\usepackage{hyperref}
\usepackage{standalone}
\usepackage{xcolor}
\usepackage{color}
\usepackage[flushleft]{threeparttable}

\usepackage{algorithm,algpseudocode}
\usepackage{subcaption}
\usepackage[makeroom]{cancel}

\newcommand{\mbR}{{\mathbb R}}
\def\bSig\mathbf{\Sigma}
\DeclareMathOperator{\vect}{vec}

\title[A Bayesian Convolutional Neural Network-based Generalized Linear Model]{A Bayesian Convolutional Neural Network-based Generalized Linear Model}

\author{Yeseul Jeon$^{1}$, 
Won Chang$^{3,4}$, 
Seonghyun Jeong$^{1,2}$,
Sanghoon Han$^{5}$,
and 
Jaewoo Park$^{1,2,*}$\email{jwpark88@yonsei.ac.kr} \\
$^{1}$Department of Statistics and Data Science, Yonsei University, Seoul, South Korea\\
$^{2}$Department of Applied Statistics, Yonsei University, Seoul, South Korea\\
$^{3}$Division of Statistics and Data Science, University of Cincinnati, Cincinnati, Ohio, U.S.A \\
$^{4}$Department of Statistics, Seoul National University, Seoul, South Korea\\
$^{5}$Department of Psychology, Yonsei University, Seoul, South Korea}

\begin{document}

\label{firstpage}


\begin{abstract}
Convolutional neural networks (CNNs) provide flexible function approximations for a wide variety of applications when the input variables are in the form of images or spatial data. Although CNNs often outperform traditional statistical models in prediction accuracy, statistical inference, such as estimating the effects of covariates and quantifying the prediction uncertainty, is not trivial due to the highly complicated model structure and overparameterization. To address this challenge, we propose a new Bayesian approach by embedding CNNs within the generalized linear models (GLMs) framework. We use extracted nodes from the last hidden layer of CNN with Monte Carlo (MC) dropout as informative covariates in GLM. This improves accuracy in prediction and regression coefficient inference, allowing for the interpretation of coefficients and uncertainty quantification. By fitting ensemble GLMs across multiple realizations from MC dropout, we can account for uncertainties in extracting the features. We apply our methods to biological and epidemiological problems, which have both high-dimensional correlated inputs and vector covariates. Specifically, we consider malaria incidence data, brain tumor image data, and fMRI data. By extracting information from correlated inputs, the proposed method can provide an interpretable Bayesian analysis. The algorithm can be broadly applicable to image regressions or correlated data analysis by enabling accurate Bayesian inference quickly. 
\end{abstract}


\maketitle

\section{Introduction}
\label{s:intro}

With recent advances in data collection technologies, it is increasingly common to have images as part of the collected data. Specifically, we consider three important applications: (1) Malaria data contain spatially correlated confirmed cases. Detecting disease hotspots and studying spatial patterns of incidence are crucial for public health problems. (2) Brain tumor data contain magnetic resonance imaging (MRI) images and texture-based statistics. Developing automated decision rules about brain tumor status can be beneficial to practitioners. (3) Anxiety disorder example contains functional magnetic resonance imaging (fMRI) data with psychological assessment collected from the survey questions. Analyzing the relationship between brain images and survey responses is an important psychological contribution. For all cases, we have both correlated matrix images and vector covariates, which are challenging to analyze simultaneously. Traditional statistical approaches face inferential issues when dealing with input image variables due to high dimensionality and multicollinearity. Convolutional neural networks (CNNs) are popular approaches for image processing by alleviating such multicollinearity issues \citep{krizhevsky2012imagenet}. However, CNNs cannot quantify uncertainties, which are often of primary interest in statistical inference. Furthermore, the complex structure of CNN can lead to nonconvex optimization, which cannot guarantee the convergence of weight parameters \citep{alzubaidi2021review}. We propose a computationally efficient Bayesian approach by embedding CNNs within generalized linear models (GLMs). Our method can take advantage of both statistical models and advanced neural networks (NNs).

There is a growing literature on Bayesian deep learning methods \citep{neal2012bayesian} to quantify the uncertainty by regarding weight parameters in deep neural networks (DNNs) as random variables. With a choice of priors, we have posterior distributions for the standard Bayesian inference. Bayesian alternatives can be robust for a small sample size compared to the standard DNN \citep{gal2016bayesian, gal2016dropout}. Despite its apparent advantages, Bayesian DNNs have been much less used than the classical one due to the difficulty in exploring the full posterior distributions. To circumvent the difficulty, variational Bayes (VB) methods have been proposed to approximate complex high-dimensional posteriors; examples include VB for DNNs \citep{tran2020bayesian} and CNNs \citep{shridhar2019comprehensive}. Although VB can alleviate computational costs, to the best of our knowledge, no existing approach provides a general Bayesian framework to study the image and vector-type variables simultaneously. Furthermore, most previous works have focused on prediction uncertainty but not on estimation uncertainty of model parameters; interpretations of regression coefficients have not been studied much in the Bayesian deep learning context.

We propose a Bayesian convolutional generalized linear model (BayesCGLM) that can study different types of variables simultaneously in various applications. First, we train Bayesian CNN (BayesCNN) \citep{gal2016bayesian} with the correlated input (images or correlated variables) and output. Then, we extract Monte Carlo (MC) samples of features from the last layer of the fitted BayesCNN. Lastly, we fit a GLM by regressing an output variable on augmented covariates with the features obtained from each MC sample. For each MC sample of the feature, we fit an individual GLM and aggregate its posterior in parallel. These steps allow us to account for uncertainties in estimating the features. We also study an often overlooked challenge in Bayesian deep learning: practical issues in obtaining credible intervals of parameters and assessing whether they provide nominal coverage.

In our method, we model BayesCNN via a deep Gaussian process (GP) \citep{damianou2013deep}, which is a nested GP. GPs \citep{rasmussen2003gaussian} are widely used to represent complex function relationships due to their flexibility. However, due to its nested functional compositions, Bayesian inference for deep GP is computationally expensive. Recently, \citet{sauer2023vecchia} approximated each layer of the deep GP through the Vecchia approximation, which scales linearly. With somewhat different perspectives, \citet{gal2016bayesian, gal2016dropout} developed
VB for deep GP based on MC approximations. 
They showed that NNs with dropout applied after every hidden layer are equivalent to the approximation of the posteriors of the deep GP. We build upon their approach to sample features from the approximate posterior of the last layer in BayesCNN.
Training BayesCNN with dropout approximation does not require additional computing costs compared to the frequentist one. There have been several recent proposals to extract features from the hidden layer of NN models \citep{tran2020bayesian,bhatnagar2020computer,fong2021forward}, and our method is motivated by them. The features can be regarded as reduced-dimensional summary statistics that measure complex non-linear relationships between the high-dimensional input and output. We observe that CNNs are specifically useful for extracting information from correlated inputs. There have been a number of recent proposals for forecasting spatio-temporal processes through ensembles of echo state networks \citep{mcdermott2017ensemble,mcdermott2019deep}. Similarly, \citet{daw2022reds} developed the random ensemble deep spatial approach that combined a deep neural model with random weights. Our work has similarities to these previous attempts in that we aggregate fitted models obtained from each MC sample of the feature. Such ensemble techniques can improve prediction over a single model and quantify uncertainties \citep{sagi2018ensemble}.

In addition to methodological improvements, our method can contribute to scientific advancement. We study the relationships between the features and response variables via principal component analysis (PCA), providing valuable insights. Furthermore, by providing adequate uncertainty quantification in both estimation and prediction, our method can aid statistically sound scientific decisions.

The outline of the remainder of this paper is as follows. In Section 2, we introduce a Bayesian DNN based on GP approximation and its estimation procedure. In Section 3, we propose an efficient BayesCGLM and provide implementation details. In Section 4, we apply BayesCGLM to several real-world data sets. We conclude with a discussion in Section 5. 

\section{A Bayesian Deep Neural Network: Gaussian Process Approximation} 
\label{BayesDeep}

Here, we provide a brief overview of Bayesian DNNs and their approximations to deep GPs. Let $\lbrace (\mathbf{x}_n,\mathbf{y}_n), n=1,\cdots, N\rbrace$ be the observed input-response pairs. A standard DNN is used to model the relationship between the input $\mathbf{x}_n$ and the output $\mathbf{o}_n \in \mbR^{d}$, which can be regarded as an expectation of $\mathbf y_n$ given $\mathbf x_n$, from a statistical viewpoint. Suppose we consider a DNN with $L$ layers, where the $l$th layer has $k_l$ nodes for $l=1,\cdots, L$. Then we can define the weight matrix $\mathbf{W}_l \in \mbR^{k_{l} \times k_{l-1}}$ that connects the $(l-1)$th layer to the $l$th hidden layer and the bias vector  $\mathbf{b}_l \in \mbR^{k_{l}}$. Then $\bm{\theta}= \lbrace (\mathbf{W}_l,\mathbf{b}_l), l=1,\cdots, L\rbrace$ is a set of parameters in the DNN that belongs to a parameter space $\bm{\Theta}$. 

To model complex data, the number of NN parameters grows tremendously with increasing $l$ and $k_l$. Such large networks are computationally intensive and can lead to overfitting issues. Dropout \citep{srivastava2014dropout} is a simple but effective strategy to address both problems. The term ``dropout" refers to randomly removing connections from nodes with some probability during training. 
The structure of NNs with dropouts is
\begin{equation}
\mathbf{o}_{n}=\sigma_L \Big(\mathbf{W}_{L}\sigma_{L-1}\Big(\cdots \sigma_3 \Big(\mathbf{W}_{3}\sigma_2 \Big(\mathbf{W}_{2} \sigma_1 \Big(\mathbf{W}_{1}\mathbf{x}_{n} +\mathbf{b}_1\Big)\circ\mathbf{d}_{2} +\mathbf{b}_2\Big)\circ\mathbf{d}_{3} +\mathbf{b}_3\Big) \cdots \Big)\circ\mathbf{d}_{L} +\mathbf{b}_{L} \Big),
\label{eq1drop}
\end{equation}
where $\sigma_l(\cdot)$ is an activation function; for example, the rectified linear (ReLU) or the hyperbolic tangent function (TanH) can be used.
Here, $\mathbf{d}_{l} \in \mbR^{k_l}$, $l=2,\cdots,L$, are dropout vectors whose elements follow a Bernoulli distribution with a pre-specified success probability $\psi_l$ independently. The notation  ``$\circ$" indicates the elementwise multiplication of vectors. As in \cite{gal2016dropout}, we apply dropout to the output of every $\sigma_{l}(\cdot)$ (i.e., the nonlinear feature). Since $\sigma_{l}(\cdot)$ has $\mathbf{W}_{l}$ and $\mathbf{b}_l$ in its argument, both weight and bias terms are regularized. Although dropout can alleviate overfitting issues, quantifying uncertainties in predictions or parameter estimates remains challenging for the frequentist approach.

To address this difficulty, Bayesian methods have been proposed by considering probabilistic alternatives of NNs. 
The Bayesian approaches consider an arbitrary function that is likely to generate data. Specifically, a deep GP has been widely used to represent complex NN structures for both supervised and unsupervised learning problems \citep{gal2016dropout, damianou2013deep}. Define $\mathbf{f}_{n,1}= \mathbf{W}_1 \mathbf{x}_n  + \mathbf{b}_1 \in \mbR^{k_1}$ and $\mathbf{f}_{n,l} = \mathbf{W}_{l} \bm{\phi}_{n,l-1} + \mathbf{b}_l \in \mbR^{k_l}$, $l\geq 2$, where $\bm{\phi}_{n,l}=\sigma_{l}(\mathbf{f}_{l}) \in \mbR^{k_l}$ is the nonlinear output feature from the $l$th layer. For Bayesian inference, we set independent normal priors for weight and bias parameters as
$p(\mathbf{W}_{l})$ and $ p(\mathbf{b}_l)$ respectively. We denote the set of linear features $\mathbf{f}_{n,l}$ as $\mathbf{F}_l = \lbrace \mathbf{f}_{n,l} \rbrace_{n=1}^{N}\in \mathbb{R}^{N\times k_l}$ and the $k$th column of $\mathbf{F}_l$ as $\mathbf{F}^{(k)}_{l}~(k=1,\cdots,k_l)$. The columns of $\mathbf{F}_{l}$ are assumed to be independent of each other for given $\mathbf{F}_{l-1}$. Then, the deep GP model can be represented as
\begin{equation}
\begin{split}
\mathbf{F}^{(k)}_{l}|\mathbf{F}_{l-1} &\sim N(0, \widehat{\mathbf{\Sigma}}_{l}),\quad l=2,\dots,L\\
      \mathbf{y}_n|\mathbf{f}_{n,L-1} &\sim p(\mathbf{y}_n|\mathbf{f}_{n,L-1}).
\label{multilayers}
\end{split}
\end{equation}
Here the empirical covariance matrix $\widehat{\mathbf{\Sigma}}_{l} \in \mbR^{N \times N}$ is
\begin{equation}
\begin{split}
    \widehat{\mathbf{\Sigma}}_{l} &= \frac{1}{k_{l}}\sigma_l(\bm\Phi_{l-1}\mathbf{W}^{\top}_{l}+ \mathbf{b}_{l})
    \sigma_l(\bm\Phi_{l-1}\mathbf{W}^{\top}_{l}+ \mathbf{b}_{l})^{\top},
    \label{covfuncsmultilayer}
\end{split}
\end{equation}
where $\bm\Phi_{l}$ are the collection of feature vectors $\lbrace \bm\phi_{n,l} \rbrace_{n=1}^{N}\in \mbR^{N\times k_l}$. This empirical covariance matrix is obtained through Monte Carlo integration, where each column of $\mathbf{W}_l$ and each element of $\mathbf{b}_l$ are sampled from the $p(\mathbf{W})$ and $p(\mathbf{b})$, respectively. According to \citet{gal2016dropout}, \eqref{covfuncsmultilayer} becomes accurate in approximating the true covariance matrix as the value of $k_l$ increases. It implies that an arbitrary data-generating function is modeled through a nested GP with the covariance of the extracted features from the previous hidden layer. In \eqref{multilayers}, $\mathbf{y}_n$ is a response variable that has a distribution  $p(\cdot|\mathbf{f}_{n,L-1})$ belonging to the exponential family with a continuously differentiable link function $g$ such that $g(E[\mathbf{y}_n|\bm{\phi}_{n,L-1}])=\mathbf{W}_{L}\bm{\phi}_{n,L-1}+\mathbf{b}_{L}$.  Note that $g^{-1}(\cdot)$ can be regarded as $\sigma_{L}(\cdot)$ in \eqref{eq1drop}. As in the typical mixed effect models, conditional independence of $\lbrace \mathbf{y}_n \rbrace_{n=1}^{N}$ is assumed.

Since Bayesian inference for deep GP using Markov chain Monte Carlo (MCMC) is not straightforward to implement, 
\cite{gal2016bayesian} developed VB for deep GP based on MC approximation, which is called {\it{MC dropout}}. 
They used a normal mixture distribution as a variational distribution $q(\bm\theta)$ to approximate the posterior distribution $\pi(\bm\theta|\lbrace\mathbf{x}_n,\mathbf{y}_n\rbrace_{n=1}^{N})$ of deep GP. Specifically, the variational distributions are defined as
\begin{equation}\begin{split}
    q(\mathbf{W}_{l}) & = \prod_{\forall i,j} q(w_{l,ij}),~~~q(\mathbf{b}_{l}) = \prod_{\forall i} q(b_{l,i})\\
    q(w_{l,ij}) &= p_lN(\mu^{w}_{l,ij},\sigma^2)+(1-p_l)N(0,\sigma^2) \\
    q(b_{l,i}) &= p_lN(\mu^{b}_{l,i},\sigma^2)+ (1-p_l)N(0,\sigma^2),
\end{split}
\label{variationaldistsuppl}
\end{equation}
where $w_{l,ij}$ is the $(i,j)$th element of the weight matrix $\mathbf{W}_{l}\in \mbR^{k_l\times k_{l-1}}$ and $b_{l,i}$ is the $i$th element of the bias vector $\mathbf{b}_{l} \in \mbR^{k_l}$. For the weight parameters, $\mu^{w}_{l,ij}$ and $\sigma^2$ are variational parameters that control the mean and spread of the distributions, respectively. As the inclusion probability $p_l \in [0,1]$ becomes close to 0, $q({w}_{l,ij})$ becomes $N(0,\sigma^2)$, indicating that it is likely to drop the weight parameters (i.e., $w_{l,ij}=0$). Similarly, the variational distribution for the bias parameters is modeled with a mixture normal distribution. Then, the optimal variational parameters are set by minimizing the Kullback–Leibler (KL) divergence between $q(\bm{\theta})$ and $\pi(\bm{\theta}|\lbrace\mathbf{x}_n,\mathbf{y}_n\rbrace_{n=1}^{N})$ where minimizing the KL divergence is equivalent to maximizing $\mbox{E}_q[{\log(\pi(\bm{\theta},\lbrace\mathbf{x}_n,\mathbf{y}_n\rbrace_{n=1}^{N}))}]-\mbox{E}_q[\log q({\bm{\theta}})]$, the evidence lower bound (ELBO). With the independent variational distribution $q(\bm{\theta}):=\prod_{l=1}^{L}q(\mathbf{W}_l)q(\mathbf{b}_l)$, the log ELBO of the deep GP is
\begin{equation} \begin{split}
     \mathcal{L}_{\text{GP-VI}} &:=  \sum_{n=1}^{N}\int \cdots \int \prod_{l=1}^{L}q(\mathbf{W}_{l})q(\mathbf{b}_{l}) \log p(\mathbf{y}_n|\mathbf{x}_n,\lbrace \mathbf{W}_{l}, \mathbf{b}_{l} \rbrace_{l=1}^{L}) d\mathbf{W}_1d\mathbf{b}_{1}\cdots d\mathbf{W}_L d\mathbf{b}_{L} \\
     &\quad - \text{KL}\Big (\prod_{l=1}^{L}q(\mathbf{W}_{l})q(\mathbf{b}_{l})\Big |\Big |p( \lbrace\mathbf{W}_{l},\mathbf{b}_{l}\rbrace_{l=1}^{L}) \Big ).
\end{split}
\label{gaussianVI}
\end{equation}
To emphasize the dependency of $\lbrace \mathbf{W}_{l}, \mathbf{b}_{l} \rbrace_{l=1}^{L}$, we represent $p(\mathbf{y}_n|\mathbf{f}_{n,L-1})$ in \eqref{multilayers} as $p(\mathbf{y}_n|\mathbf{x}_n,\lbrace \mathbf{W}_{l}, \mathbf{b}_{l} \rbrace_{l=1}^{L})$, which is identical. Since the direct maximization of \eqref{gaussianVI} is challenging due to the intractable integration, \citet{gal2016dropout} replaced it with MC approximation as 
\begin{equation}
\begin{split}
\mathcal{L}_{\text{GP-MC}} & :=\frac{1}{M}\sum_{m=1}^{M}\sum_{n=1}^{N} \log p(\mathbf{y}_n|\mathbf{x}_n,\lbrace \mathbf{W}_{l}^{(m)}, \mathbf{b}_{l}^{(m)} \rbrace_{l=1}^{L}) \\
&\quad -\text{KL}\Big(\prod_{l=1}^{L}q(\mathbf{W}_{l})q(\mathbf{b}_{l})\Big|\Big|\prod_{l=1}^{L}p(\mathbf{W}_{l})p(\mathbf{b}_{l})\Big),
\label{GPMCsuppl}
\end{split}
\end{equation} 
where $\lbrace\lbrace \mathbf{W}_{l}^{(m)}, \mathbf{b}_{l}^{(m)} \rbrace_{l=1}^{L}\rbrace_{m=1}^{M}$ is MC samples from the variational distribution in \eqref{variationaldistsuppl}.  

Note that the MC samples are generated for each stochastic gradient descent (SGD) update, and the estimates from $\mathcal{L}_{\text{GP-MC}}$ would converge to those obtained from $\mathcal{L}_{\text{GP-VI}}$ \citep{paisley2012variational,rezende2014stochastic}. Furthermore, \cite{gal2016dropout} showed that \eqref{GPMCsuppl} converges to the frequentist loss function of a DNN with dropouts when $\sigma$ in \eqref{variationaldistsuppl} is small value and $k_l$ becomes large. This result implies that the standard DNNs with dropout applied after every hidden layer are equivalent to the approximation of the posteriors of the deep GP. We provide mathematical details in Web Appendix A.1.

In many applications, the prediction of unobserved responses is of great interest. Consider an unobserved response $\mathbf{y}_{n}^{\ast}\in \mbR^{d}$ with an input $\mathbf{x}_{n}^{\ast} \in \mbR^{k_0}$. For given $\mathbf{x}_{n}^{\ast}$, we can obtain $\bm{\phi}^{\ast}_{n,l} \in \mbR^{k_l}$ from the approximate posterior predictive distribution based on MC dropout. Although \citet{gal2016dropout} focused on predicting $\mathbf{y}_{n}^*$, this procedure allows us to construct informative and low-dimensional summary statistics for $\mathbf{x}_{n}^{\ast}$. Specifically, we can sample an output feature vector $\bm{\phi}^{\ast}_{n, L-1} \in \mbR^{k_{L-1}}$ which summarizes the complex nonlinear dependence relationships between input and output variables up to the last layer of the NN. In Section 3, we use these extracted features as additional covariates in the GLM framework.

\section{Flexible Generalized Linear Models with Convolutional Neural Networks}

As a variant of DNNs, CNNs have been widely used in computer vision and image classification \citep{krizhevsky2012imagenet, shridhar2019comprehensive}. CNNs can effectively extract important features from images or correlated datasets with nonlinear dependencies. In our scientific applications, we have spatially correlated datasets for malaria incidence, MRI images for brain tumors, and a fMRI correlation matrix of the brain for anxiety disorder. Given that our applications involve image or correlated datasets, CNNs are well-suited for capturing structural dependency by extracting local information through convolution layers.

We propose a Bayesian convolutional generalized linear model (BayesCGLM) based on two appealing methodologies: (1) CNNs have been widely used to utilize correlated predictors (or input variables) with spatial structures such as images or geospatial data. (2) GLMs can study the relationship between covariates and non-Gaussian responses by extending a linear regression framework. Let $\mathbf{D}=\lbrace \mathbf{X},\mathbf{Y}, \mathbf{Z} \rbrace$ be the observed dataset, where $\mathbf{X}=\lbrace \mathbf{X}_n \rbrace_{n=1}^{N}$ be the collection of correlated input (i.e., each $\mathbf{X}_n$ is a matrix or tensor) and  $\mathbf{Y}=\lbrace \mathbf{y}_n \rbrace_{n=1}^{N}$ be the corresponding output. We also let $\mathbf{Z}=\lbrace \mathbf{z}_n \rbrace_{n=1}^{N}$ be a collection of $p$-dimensional vector covariates. We use BayesCNN to extract the feature from $\mathbf{X}$ and employ it in a GLM with the covariates $\mathbf{Z}$ to predict $\mathbf{Y}$.

Such an approach is useful in that the feature is optimized for the prediction of $\mathbf{Y}$. Since the feature is guided by $\mathbf{Y}$, we can construct a map from the features to the output. Similar to our approach, \citet{bhatnagar2020computer} also extracted features through DNN and used the features to fit their model. They showed that this can improve the prediction performance of the model. In Section 4, we observe that the feature serves as an important summary statistic for predicting response $\mathbf{Y}$. Compared to \cite{bhatnagar2020computer}, we also consider the uncertainties in estimating the features using MC dropout.

\subsection{Bayesian Convolutional Neural Network}\label{Sec:BayesCNN}

A Bayesian perspective can quantify uncertainties through posterior densities of $\bm{\theta}$. Specifically, \citet{gal2016bayesian} proposed an efficient Bayesian alternative by regarding weight parameters in CNN's kernels as random variables. By extending a result in \citet{gal2016dropout}, \citet{gal2016bayesian} showed that applying dropout after every convolution layer can approximate the intractable posterior distribution of deep GP.

Let $\mathbf{X}_n\in \mbR^{M_0 \times R_0 \times C_0}$ be an input with $M_0$ height, $R_0$ width, and $C_0$ channels. For example, $M_0$ and $R_0$ are determined based on the resolution of images, and $C_0=3$ for color images ($C_0=1$ for black and white images). There are three types of layers in CNN: the convolution, pooling, and fully connected layers. Convolution layers move kernels (filters) on the input image and extract information, resulting in a ``feature map".  Each cell in the feature map is obtained from element-wise multiplication between the image and the filter matrix. The weight values in filter matrices are parameters to be estimated. Higher output values indicate stronger signals on the image, which can account for spatial structures. We can extract the important neighborhood features from the input by shifting the kernels over all pixel locations with a certain step size (stride). Then, the pooling layers are applied to downsample the feature maps. The convolution and pooling layers are applied repeatedly to reduce the dimension of an input image. Then, the extracted features from the convolution/pooling layers are flattened and connected to the output layer through a fully connected layer, which can learn non-linear relationships between the features and the output.

Consider a CNN with $L$ number of layers, where the last two are fully connected and output layers. For the $l$th convolution layer, there are $C_l$ number of kernels $\mathbf{K}_{l,c} \in \mbR^{H_{l} \times D_{l} \times C_{l-1}}$ with $H_l$ height, $D_l$ width, and $C_{l-1}$ channels for $c=1,\cdots,C_l$. Note that the dimension of the channel is determined by the number of kernels in the previous convolution layer. After applying the $l$th convolution layer, the $(m,r)$th element of the feature matrix $\bm{\eta}_{n,(l,c)}\in \mbR^{M_{l} \times R_{l}}$ becomes
\begin{equation}
     [\bm{\eta}_{n,(l,c)}]_{m,r}= \sigma \Big(\sum_{i=1}^{H_{l}}\sum_{j=1}^{D_{l}}\sum_{k=1}^{C_{l-1}} ([\mathbf{K}_{l,c}]_{i,j,k} [\bm{\eta}_{n,(l-1,c)}]_{m+i-1,r+j-1,k}) + b_{l,c}\Big),
\label{convolutionweight}
\end{equation}
where $b_{l,c}$ is the bias parameter. For the first convolution layer, $\bm{\eta}_{n,(l-1,c)}$ is replaced with an input image $\mathbf{X}_n$. (i.e., $\bm{\eta}_{n,(0,c)}=\mathbf{X}_n$). For computational efficiency and stability, ReLU functions are widely used as activation functions $\sigma(\cdot)$. By shifting kernels, the convolution layer can capture neighborhood information, and the neighborhood structure is determined by the size of the kernels. \citet{gal2016bayesian} pointed out that extracting the features from the convolution operation can be reformulated as an affine transformation in standard DNN (details are in Web Appendix A.2). Therefore, CNNs can also be represented as a deep GP with dropout approximation; applying dropout after every convolution layer before pooling can approximate a deep GP. 

Note that optimal hyperparameter tuning in NNs requires a number of empirical fittings by varying hyperparameters. \cite{wang2019jumpout} pointed out that a higher dropout rate can lead to a slower convergence rate, while a lower rate can deteriorate generalization performance. Although \cite{gal2016dropout, gal2016bayesian} used a dropout rate of around 0.1-0.5, it does not always guarantee optimal performance. Therefore, we recommend conducting empirical studies by varying dropout rates within such a range for validation. In Web Appendix B, we observe that our method is robust across different dropout rates.

After the $L-2$th convolution layers, the extracted feature tensor $\bm\eta_{n,L-2} \in \mbR^{M_{L-2} \times R_{L-2} \times C_{L-2}}$ is flattened in the fully connected layer. By vectorizing $\bm\eta_{n,L-2}$, we can represent the feature tensor as a feature vector $\bm\phi_{n,L-2}=\vect(\bm\eta_{n,L-2}) \in \mbR^{k_{L-2}}$, where $k_{L-2} = M_{L-2} R_{L-2} C_{L-2}$. Finally, we can extract a feature vector $\bm\phi_{n,L-1} \in \mbR^{k_{L-1}}$ from the fully connected layer. From here, we define the collection of the extracted feature vector as $\bm{\Phi}=\lbrace \bm\phi_{n, L-1} \rbrace_{n=1}^{N}$. We suppress the index $L-1$ in $\bm{\Phi}_{L-1}$ for notational convenience.

The training step of BayesCNN is identical to that of standard CNN. In the prediction step, the algorithm performs forward propagation for the given estimated NN parameters. For each iteration, $\lbrace \mathbf{d}_l \rbrace_{l=2}^{L}$ are randomly sampled from the Bernoulli distribution with rate $\psi_l$ and are applied to the features. The algorithm repeats this $M$ times to obtain an MC sample of $M$ predicted outputs. We provide implementation details for BayesCNN in Web Appendix C.1.

\subsection{Bayesian Convolutional Generalized Linear Model}

We propose BayesCGLM by regressing $\mathbf{Y}$ on $\bm{\Phi}$ and additional covariates $\mathbf{Z}$. Here, we use $\bm{\Phi}$ as a basis matrix that encapsulates information of high-dimensional and correlated input variables $\mathbf{X}$. Let $\mathbf{A}=[\mathbf{Z}, \bm{\Phi}] \in \mbR^{N \times (p+k_{L-1})}$ be the design matrix. Then, our model is
\begin{equation} \begin{split}g(E[\mathbf{Y}|\mathbf{Z},\bm{\Phi}]) &= \mathbf{Z}\bm{\gamma} + \bm{\Phi}\bm{\delta} = \mathbf{A}\bm{\beta},
\end{split}
\label{model}
\end{equation}
where $\bm{\beta}=(\bm{\gamma}^{\top}, \bm{\delta}^{\top})^{\top} \in \mbR^{p+k_{L-1}}$ is the corresponding regression coefficients and $g(\cdot)$ is a one-to-one continuously differential link function. The proposed method can provide an interpretation of the regression coefficients and quantify uncertainties in prediction and estimation. Furthermore, the training step does not require additional computational costs compared to the traditional CNN. We train BayesCNN with the negative log-likelihood loss function corresponding to the type of output variable. To be more specific, we used the mean squared loss, binary cross-entropy loss, and Poisson loss for Gaussian, binary, and count outputs, respectively. Such choices are natural in that the loss functions are directly defined by the distribution of the output variable. 

Since we place a posterior distribution over the kernels of the fitted BayesCNN (Section~\ref{Sec:BayesCNN}), we can generate MC samples of the features $\bm{\Phi}^{(m)}$, $m=1,\cdots,M$ from the variational distribution. To account for the uncertainties in estimating the features, we use all the MC samples of the features rather than their point estimates. For each MC sample of $\mathbf{A}^{(m)}=[\mathbf{Z}, \bm{\Phi}^{(m)}]$, we can fit individual GLM in parallel; we have $ M$ number of posterior distributions of $\bm{\beta}_m=(\bm{\gamma}_m^{\top}, \bm{\delta}_m^{\top})^{\top} \in \mbR^{p+k_{L-1}}$. This step is parallelizable, so computational wall times decrease with more cores. Then, we construct an aggregated distribution (the ``ensemble-posterior") of $\bm{\beta}$. In the following section, we provide details about constructing the ensemble-posterior distribution. The fitting procedure of BayesCGLM is summarized in Web Appendix C.2.

In general, inference for the regression coefficients $\bm{\gamma}$ in \eqref{model} is challenging for DNN. Although one can regard the weight parameters at the last hidden layer as $\bm{\gamma}$, we cannot obtain uncertainties in estimates. Even BayesDNN \citep{gal2016dropout} only provided point estimates of weight parameters, while they can quantify uncertainties in response prediction. On the other hand, our method directly provides the posterior distribution of $\bm{\gamma}$, which is useful for standard inference. In addition, the complex structure of DNN often leads to nonconvex optimization \citep{goodfellow2014qualitatively,dauphin2014identifying}, resulting in an inaccurate estimate of $\bm{\gamma}$. Since the stopping rule of the SGD algorithm is based on prediction accuracy (not on convergence in parameter estimation), the convergence of individual parameters, including $\bm{\gamma}$ cannot be guaranteed. The problem gets further complicated by a large number of parameters that need to be simultaneously updated in the SGD algorithm. Note that BayesCGLM also uses the SGD algorithm to extract $\bm{\Phi}$. From this, we can obtain one of the optimal combinations of weight parameters and the corresponding $\bm{\Phi}$. Although such $\bm{\Phi}$ may not be the global optimizer, it is still informative to predict responses because it is obtained by minimizing the loss function. However, this does not guarantee the convergence of individual weight parameters in nonconvex optimization. Instead, our two-stage approach can alleviate this issue by using extracted $\bm{\Phi}$ as a fixed basis matrix in GLM. In Web Appendix D, we conduct the simulation studies under different model configurations and show that BayesCGLM can recover the true $\bm{\gamma}$ values well, while estimates from BayesCNN can be biased.

\subsection{Bayesian Inference}

Here, we describe a Bayesian inference for the proposed method. We first define the posterior distribution of the regression coefficient $\pi(\bm\beta|\mathbf{D})$. Let $\lbrace \mathbf{W}_{l},\mathbf{b}_{l} \rbrace_{l=1}^{L-1}$ be the weights and biases corresponding to all hidden layers in CNN. We can vectorize kernels in CNN and represent them as weights as in the standard DNN (see Web Appendix A.2 for details). Note that $\bm\Phi$ can be deterministically obtained from forward propagation for the given $\mathbf{X}$ and $\lbrace \mathbf{W}_{l},\mathbf{b}_{l} \rbrace_{l=1}^{L-1}$. Therefore, the posterior distribution can be represented as
\begin{equation} \begin{split}
     \pi(\bm\beta|\mathbf{D}) &= \int \pi(\bm\beta|\mathbf{D},\lbrace \mathbf{W}_l,\mathbf{b}_l \rbrace_{l=1}^{L-1}) \\
     & \quad  \times \prod_{l=1}^{L-1} \pi(\mathbf{W}_l|\mathbf{D}) \pi(\mathbf{b}_l|\mathbf{D}) d\mathbf{W}_1 d\mathbf{b}_1 \cdots d\mathbf{W}_{L-1} d\mathbf{b}_{L-1}, 
\end{split}
\label{betapost}
\end{equation}
where, $\pi(\bm\beta| \mathbf{D}, \lbrace \mathbf{W}_l,\mathbf{b}_l \rbrace_{l=1}^{L-1})$ is the conditional posterior, and $\pi(\mathbf{W}_l|\mathbf{D})$, $\pi(\mathbf{b}_l|\mathbf{D})$ are marginal posteriors for weight and bias, respectively. Since it is challenging to compute \eqref{betapost} directly, we approximate it through MC dropout as
\begin{equation} 
       \int \pi(\bm\beta| \mathbf{D}, \lbrace \mathbf{W}_l,\mathbf{b}_l \rbrace_{l=1}^{L-1}) \prod_{l=1}^{L-1} q(\mathbf{W}_l) q(\mathbf{b}_l) d\mathbf{W}_1 d\mathbf{b}_1 \cdots  d\mathbf{W}_{L-1} d\mathbf{b}_{L-1}, 
\label{beta2}
\end{equation}
where $q(\mathbf{W}_l)$ and $q(\mathbf{b}_l)$ are variational distributions defined in \eqref{variationaldistsuppl}. Then the MC approximation to \eqref{beta2} is
\begin{equation} 
\frac{1}{M}\sum_{m=1}^{M} \pi(\bm\beta_m|\mathbf{D},\lbrace \mathbf{W}^{(m)}_{l},\mathbf{b}^{(m)}_{l} \rbrace_{l=1}^{L-1}).
\label{MCapproxRES}
\end{equation}
Here $\lbrace \lbrace \mathbf{W}^{(m)}_{l},\mathbf{b}^{(m)}_{l} \rbrace_{l=1}^{L-1}\rbrace_{m=1}^{M}$ are sampled from \eqref{variationaldistsuppl}.  
One may choose the standard MCMC algorithm to estimate each $ \pi(\bm\beta_m|\mathbf{D},\lbrace \mathbf{W}^{(m)}_{l},\mathbf{b}^{(m)}_{l} \rbrace_{l=1}^{L-1})$ in \eqref{MCapproxRES}. However, this may require a long run of the chain to guarantee good mixing, resulting in excessive computational cost. Instead, we use the Laplace method to approximate each $\pi(\bm\beta_m|\mathbf{D},\lbrace \mathbf{W}^{(m)}_{l},\mathbf{b}^{(m)}_{l} \rbrace_{l=1}^{L-1})$. In our preliminary studies, we observed that the Laplace approximation leads to a similar inference result as the MCMC method within a much shorter computing time when the sample size is large enough.

For the Laplace approximation, 
we first compute $\bm\Phi^{(m)}$ for the given $\lbrace \mathbf{W}^{(m)}_{l},\mathbf{b}^{(m)}_{l} \rbrace_{l=1}^{L-1}$ through forward propagation. Then, we obtain the maximum likelihood estimate (MLE) $\widehat{\bm{\beta}}_{m}$ using GLM by regressing $\mathbf{Y}$ on $\bm\Phi^{(m)}$ and $\mathbf{Z}$. We approximate the posterior of $\bm{\beta}_{m}$ as $\mathcal{N}(\widehat{\bm{\beta}}_{m}, \widehat{\mathbf{B}}^{-1}_{m})$, where $\widehat{\mathbf{B}}_{m} \in \mbR^{(p+k_{L-1}) \times (p+k_{L-1})}$ is the observed Fisher information matrix from the $m$th MC samples. 
From this procedure, \eqref{MCapproxRES} can be approximated as
\begin{equation}
      \frac{1}{M}\sum_{m=1}^{M}  \varphi(\bm{\beta}; \widehat{\bm{\beta}}_{m},\widehat{\mathbf{B}}^{-1}_{m}),
\label{posterirdist}
\end{equation}
where $\varphi(\bm{x}; \bm{\mu},\bm{\Sigma})$ is a multivariate normal density with mean $\bm{\mu}$ and covariance $\bm{\Sigma}$. In summary, \eqref{posterirdist} is the approximation of the posterior distribution $\pi(\bm\beta|\mathbf{D})$ through MC dropout and the Laplace method. Note that BayesCGLMs approximate $\pi(\bm\beta|\mathbf{D})$ through the MC average from the same probability model (i.e., data generating mechanism). Therefore, the interpretation of $\bm{\beta}$ does not change from ensemble to ensemble. However, the interpretation of $\bm\delta$ can be more difficult because each column of $\bm{\Phi}$ comes from complex (and repeated) convolution operations. It is challenging to explain the impact of changes in features based on the estimated $\bm\delta$.

Similar to BayesCNN, our method can quantify uncertainties in predictions, which is an advantage over deterministic NNs. Note that BayesCGLM can also provide interpretations of model parameters with uncertainty quantification, while BayesCNN cannot. For unobserved response $\mathbf{Y}^{\ast} \in \mbR^{N_{\text{test}}}$, we have $\mathbf{A}^{\ast}_{m}=[\mathbf{Z}^{\ast}, \bm{\Phi}^{\ast(m)}]\in \mbR^{N_{\text{test}}\times (p+k_{L-1})}$ for $m=1,\cdots,M$. 
Here, $\bm{\Phi}^{\ast(m)}$ is $m$th features extracted from the last hidden layer of the trained BayesCNN for given $\mathbf{X}^{\ast}$ and  $\mathbf{Z}^{\ast}$. Then, from \eqref{posterirdist}, the predictive distribution of the linear predictor is 
\begin{equation}
\frac{1}{M}\sum_{m=1}^{M} \varphi(\mathbf{A}^{\ast}\bm{\beta}; \mathbf{A}_{m}^{\ast}\widehat{\bm{\beta}}_{m},\mathbf{A}_{m}^{\ast}\widehat{\mathbf{B}}^{-1}_{m}\mathbf{A}^{\ast\top}_{m}).
\label{linearpredictor}
\end{equation}
For given \eqref{linearpredictor}, we can generate $\mathbf{Y}^{\ast}$ from the underlying distribution. For example, when we have a Gaussian response, we have $\mathbf{Y}^{\ast} \sim \mathcal{N}(  \frac{1}{M}\sum_{m=1}^{M}\mathbf{A}^{\ast}\bm{\widehat{\beta}}_m,\widehat{\sigma}^2)$ where $\widehat{\sigma}^2 = \sum_{m=1}^{M}(\mathbf{A}^{(m)}\bm{\widehat{\beta}}_{m}-\mathbf{Y})^{\top}(\mathbf{A}^{(m)}\bm{\widehat{\beta}}_{m}-\mathbf{Y})/NM$. For the count response, we can simulate $\mathbf{Y}^{\ast}$ from the Poisson distribution with an intensity of $\exp(\frac{1}{M}\sum_{m=1}^{M}\mathbf{A}^{\ast}\bm{\widehat{\beta}}_m)$.

As we described, BayesCGLM involves several approximation steps: (1) MC dropout and (2) the Laplace method; therefore, investigating approximation quality is crucial. In  Web Appendix D.5, we conduct a numerical study under the simple NN setting. Here, we use a standard NN to extract features; therefore, we refer to our method as BayesDGLM. We compare the performance of BayesDGLM, BayesDNN, and a fully Bayesian method (MCMC). We observe that the BayesDGLM provides comparable inference results to the fully Bayesian method with much smaller computational costs. Furthermore, BayesDNN cannot quantify uncertainties in parameter estimates, while BayesDGLM can. Considering that it is infeasible to apply the fully Bayesian method for complex DNNs, BayesDGLM would be a practical option in many applications. We provide details for the experiment in Web Appendix D.5.

\section{Applications}\label{applications}

We apply our method to brain tumor image data and fMRI data for anxiety scores. We also provide results for geospatial application in Web Appendix E. We compared our model with BayesCNN and GLM to assess its performance. We evaluated prediction accuracy using the root mean square prediction error (RMSPE) and uncertainty quantification performance using the empirical coverage of prediction intervals. Details about the CNN structure are provided in Web Appendix F.

\subsection{Brain Tumor Image Data}

According to \citet{noone2020seer}, brain and nervous system tumors have low 5-year relative survival rates, which have gradually increased from 1975 to 2016, reaching 33\%. Furthermore, some brain tumors can also be cancerous. Considering that the 5-year survival rate for overall cancers is 67\%, brain tumors can be life-threatening. MRI is the standard diagnostic tool for brain tumors, and texture-based first- and second-order statistics of MRI images are useful for classification \citep{aggarwal2012first}, and these statistics are widely used in the classification of biomedical images \citep{ismael2018brain}. However, these statistics do not capture the spatial information of correlated structures in MRI images. Automated methods for brain tumor classification are becoming increasingly important, as manual analysis of MRI data by medical experts is challenging. It is also crucial to convey the uncertainty associated with these automated methods to ensure statistically sound decision-making. Compared to previous studies \citep{aggarwal2012first,ismael2018brain}, we use both MRI images and texture-based statistics to predict brain tumors while quantifying uncertainties. The dataset is collected from the Brain Tumor Image Segmentation Challenge \citep{menze2014multimodal}. The binary response indicates whether a patient has a fatal brain tumor. For each patient, we have an MRI image with its two summary variables (first and second-order features of MRI images). Here, we use $N = 2,508$ observations to fit our model and reserve $N_{\text{test}} = 1,254$ observations for validation. We fit BayesCNN using $240 \times 240$ pixel gray images $\mathbf{X}$, scalar covariates $\mathbf{Z}$ as input, and the binary response $\mathbf{Y}$ as output. From the fitted BayesCNN, we can extract $\bm\Phi \in \mbR^{2,508 \times 16}$ from the last hidden layer. In this study, we use two summary variables (first and second-order features of MRI images) as covariates. With the extracted features from the BayesCNN, we fit logistic GLMs by regressing $\mathbf{Y}$ on $[\mathbf{Z},\bm\Phi^{(m)}] \in \mbR^{2,508 \times 18}$ for $m=1,\cdots, 500$.

Table~\ref{table_tumor} shows inference results from different methods. Our method shows that the first-order feature has a negative relationship, while the second-order feature has a positive relationship with the brain tumor risk. Both coefficients are statistically significant based on the highest posterior density (HPD) intervals. Note that previous work \citep{aggarwal2012first} only predicted tumor status but did not provide such an interpretation from the first and second-order statistics. Therefore, we can conclude that darker MRI images (larger first-order statistics) with lower variation (smaller second-order statistics) indicate a lower possibility of tumor presence. We observe that the sign of $\bm{\gamma}$ estimates from BayesCGLM is aligned with those from GLM.

\begin{table}
\centering
\caption[]{Inference results for the brain tumor dataset from different methods. For all methods, the posterior mean of $\bm{\gamma}$, 95\% HPD interval, accuracy, recall, precision, and computing time (min) are reported in the table.}
\label{table_tumor}
\begin{tabular}{l l c c c}
& & \textbf{BayesCGLM} &  \textbf{BayesCNN} &\textbf{GLM} \\ 
& & $M=500$ & $M=500$ & \\
\hline 
$\gamma_1$ (first-order feature) & Mean  & $-5.332$ & 0.248 & $-2.591$  \\
    & 95\% Interval  & $(-7.049,-3.704)$ & - & $(-2.769,-2.412)$ \\
$\gamma_2$ (second-order feature) & Mean & 4.894 & 0.160 & 2.950 \\
    & 95\% Interval & $(3.303, 6.564)$ & - & $(2.755, 3.144)$ \\
Prediction & Accuracy & 0.924 & 0.867 & 0.784   \\    
& Recall & 0.929 & 0.787 & 0.783 \\
& Precision & 0.901  & 0.907 &  0.715  \\
Time (min)& & 293.533 & 103.924 & 0.004\\
 \hline
\end{tabular}
\end{table}

\begin{figure}[htbp]
\begin{center}
\includegraphics[width = 1\textwidth,trim={1.5mm 1.5mm 1.5mm 1.5mm}]{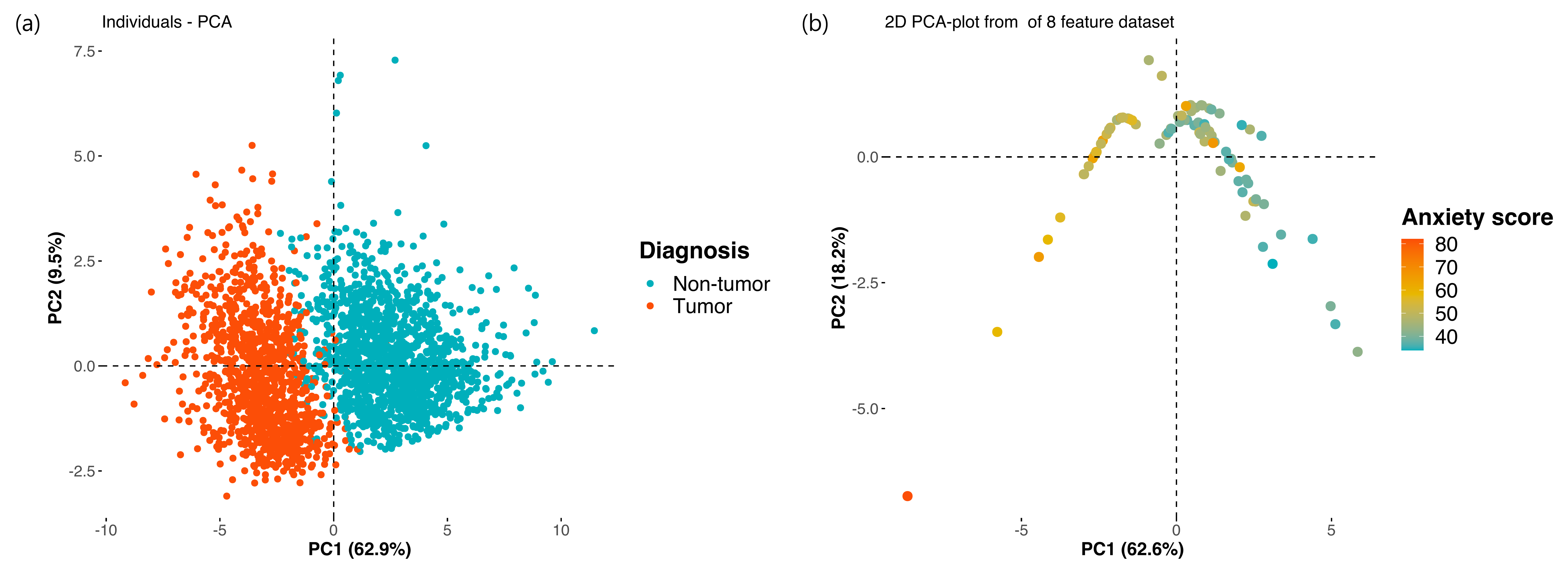}
\end{center}
\caption[]{Score plots for each application: (a) brain tumor dataset and (b) fMRI dataset.}
\label{Phi_realdata}
\end{figure}

Figure~\ref{Phi_realdata} is a score plot with the first and second principal components of $\bm\Phi$. Figure~\ref{Phi_realdata} (a) shows a clear separation of tumor and non-tumor status based on two principal components. In particular, we observe that patients without tumors mostly have positive values of the first principal component (62.9\% explained variability). This indicates that $\bm{\Phi}$ contains useful information from images for predicting brain tumor status. By incorporating $\bm{\Phi}$ information, BayesCGLM shows the most accurate prediction performance while providing credible intervals for the estimates.

\begin{figure}[htbp]
\begin{center}
\includegraphics[width = 1\textwidth]{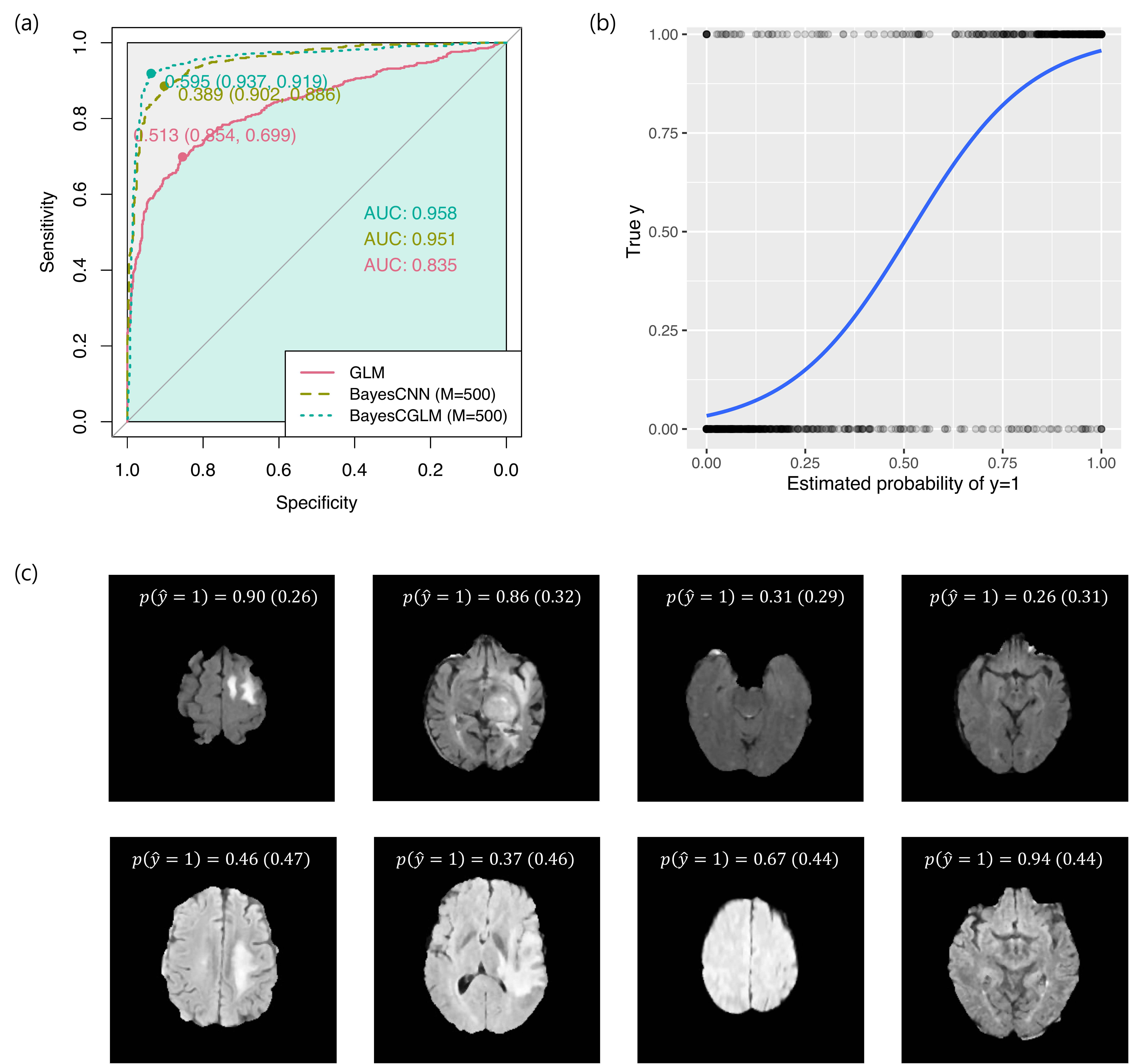}
\end{center}
\caption[]{(a) The ROC curve and AUC obtained by each method. (b) The estimated probability of $y=1$ (blue solid line) and the observed binary responses (black dots). (c) MRI images with the fitted probabilities (prediction errors) for different cases. The top panel illustrates correctly specified images with small prediction errors. The bottom panel illustrates misclassified images with large prediction errors.}
\label{tumor}
\end{figure}

 Figure~\ref{tumor} (a) shows that the area under the curve (AUC) for the receiver operating characteristic curve (ROC) plot is about 0.96, which shows outstanding prediction performance. The predicted probability surface becomes larger for the binary responses with a value of 1 (Figure~\ref{tumor} (b)). Furthermore, the true and predicted binary responses are well aligned. We also investigate how prediction uncertainties are related to actual images. The top panel in Figure~\ref{tumor} (c) shows four examples of correctly classified cases with low prediction uncertainties. The results show that when the image is overall dark, and the tumor area clearly stands out from the background, we have low uncertainties. On the other hand, the bottom panel in Figure~\ref{tumor} (c) shows four examples of misclassified cases with high prediction uncertainties. In these cases, the background image is too bright, so the tumor area is hardly distinguishable from the background. Although the point predictions misclassify these cases, higher uncertainty informs us that further investigation is needed. Therefore, better medical decisions can be made when uncertainty measures are considered.

\subsection{fMRI Data for Anxiety Score}

Anxiety disorders are becoming a prevalent mental health issue worldwide, with studies showing a more than 25\% increase in cases in 2020 \citep{world2022mental}. There have been several works to study the relationship between anxiety scores and psychological assessments collected from the survey data \citep{Kaplan2015}. However, no previous studies considered the fMRI data, which contains important information about the functional connectivity of the brain. Here, we utilize fMRI information through the extracted feature to construct an anxiety score model. From this, we can incorporate the correlation structure of the brain when we study the relationships between psychological assessments and anxiety levels.

The resting-state fMRI data is collected by the enhanced Nathan Kline Institute-Rockland Sample (NKI-RS) to create a large-scale community sample of participants across the lifespan. For 156 patients, resting-state fMRI signals were scanned for 394 seconds. Then we transform fMRI images into 278 functional brain Regions-of-Interest (ROIs) following the parcellation scheme of \citet{shen2013groupwise}. 

A functional network was computed using the pairwise Pearson correlation coefficient, constructing $278\times278$ square symmetric matrix, which is called the ROI-signal matrix. The correlation matrix of brain functions has been widely used to study the functional connectivity of the brain \citep{sporns2002network}. Therefore, we also use the correlation matrix as input $\mathbf{X}$. Participants in the study completed a survey that included various physiological and psychological assessments. In this study, we use $N=156$ observations to fit our model and reserve $N_{\text{test}}=52$ observations for validation. We use conscientiousness, extraversion, agreeableness, and neuroticism as $\mathbf{Z}$ and the averaged anxiety score as $\mathbf{Y}$. We train BayesCNN with $\mathbf{X}$, $\mathbf{Z}$ as input, and the continuous response $\mathbf{Y}$ as output. From the trained BayesCNN, we extract each $m$th feature matrix $\bm\Phi^{(m)} \in \mbR^{104 \times 8}$ from the last hidden layer. Then we regress $\mathbf{Y}$ on $[\mathbf{Z},\bm\Phi^{(m)}]$ for $m=1,\cdots,500$. 

Table~\ref{NKI_table} indicates that BayesCGLM has comparable prediction performance compared to BayesCNN and can quantify uncertainties of regression coefficients. While the sign of the estimated $\bm{\gamma}$ are identical across different methods, there are differences in the statistical significance, particularly in agreeableness and conscientiousness variables. From BayesCGLM, we can infer that the person who has emotional instability and vulnerability to unpleasant emotions (neuroticism) and who is less likely to rebel against others (agreeableness) has a high anxiety and depression score, which are aligned with the previous study in \citet{Kaplan2015}. On the other hand, individuals who are more sociable (extroverted) and driven to achieve their goals (conscientiousness) tend to exhibit lower anxiety and depression scores, as also supported by the previous studies \citep{naragon2014interaction,fan2020relationships}. Note that these previous studies neither studied the psychological assessments simultaneously nor fMRI information.

GLM, which does not utilize fMRI information, draws different conclusions in a significance test. Table~\ref{NKI_table} shows that agreeableness and conscientiousness variables are not statistically significant in GLM, which contradicts the findings in the previous studies \citep{Kaplan2015,fan2020relationships}. Figure~\ref{Phi_realdata} is a score plot with the first and second principal components of $\bm\Phi$. Figure~\ref{Phi_realdata} (b) indicates a negative correlation between anxiety scores and the first principal component (62.6\% explained variability). Therefore, $\bm\Phi$ encapsulates information about brain functions that are highly associated with anxiety levels. By incorporating such important information into the model, BayesCGLM can accurately infer the relationship between psychological assessments and show improved prediction performance.

We observe that all the methods provide similar coefficient estimates. We can infer that the person who has emotional instability and vulnerability to unpleasant emotions (neuroticism) and who is less likely to rebel against others (agreeableness) has a high anxiety and depression score, which gives the comparable result from \citet{Kaplan2015}. The person who is more inclined to seek sociability and is eager to attain goals (extraversion), on the other hand, has lower anxiety and depression scores. Note that the previous work \citep{Kaplan2015} only conducted exploratory data analysis based on correlation coefficients without quantifying the uncertainty. On the other hand, our model can estimate the coefficients by incorporating correlated fMRI information.

\begin{table}[tt]
\centering
\caption[]{Inference results for the fMRI dataset from different methods. For all methods, posterior mean of $\bm{\gamma}$, 95\% HPD interval, RMSPE, prediction coverage, and computing time (min) are reported in the table.}
\label{NKI_table}
\begin{tabular}{l l c c c}
& & \textbf{BayesCGLM} &  \textbf{BayesCNN} &\textbf{GLM} \\ 
& & $M=500$ & $M=500$ & \\
\hline
$\gamma_1$ (neuroticism) & Mean  &  3.496 &   3.480    & 4.318 \\
   & 95\% Interval & $(2.629, 4.29)$ &  - & $(2.572, 6.064)$\\
$\gamma_2$ (extraversion) & Mean &  $-1.123$ & $-1.149$   & $-1.653$\\
    & 95\% Interval & $(-1.829, -0.386)$ & - & $(-3.198, -0.108)$ \\
$\gamma_3$ (agreeableness)& Mean & 1.113 & 1.204 & 1.196\\
     & 95\% Interval      & $(0.480, 1.760)$& -  & $(-0.229, 2.621) $ \\
$\gamma_4$ (conscientiousness) & Mean  & $-1.394$ & $-1.416$ &$-1.283$\\
       & 95\% Interval & $(-2.226, -0.507)$& - & $(-3.056, 0.491)$ \\
Prediction & RMSPE &  9.342  &  9.794&  10.531 \\ 
 & Coverage & 0.923 & 0.981 & 0.923 \\
Time (min) & & 3.459 & 0.491 &0.001 \\
\hline
\end{tabular}
\end{table}

\begin{figure}[htbp]
\begin{center}
\includegraphics[width = 0.7\textwidth]{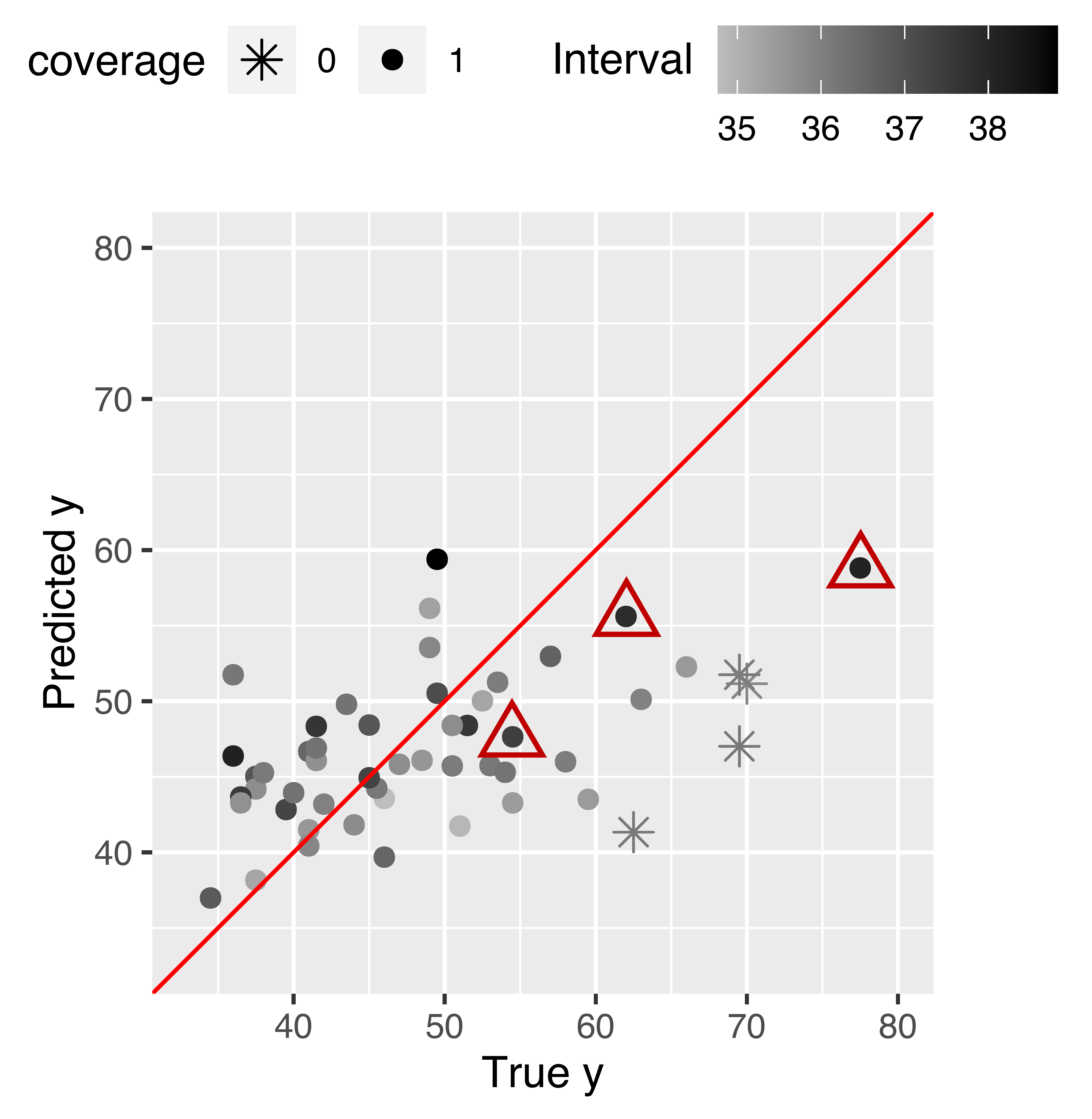}
\end{center}
\caption[]{Comparison between the true and predicted anxiety scores from BayesCGLM. Circles (asterisks) represent that the true scores are (not) covered in 95\% HPD intervals. Triangles represent individuals with the three largest HPD intervals.}
\label{NKI_results1}
\end{figure}

Figure~\ref{NKI_results1} shows that there is some agreement between the true and predicted responses. We observe that HPD prediction intervals include the true response well (92.3\% prediction coverage). Due to the small sample size, prediction intervals are wider than previous examples, which is natural.
In Figure~\ref{NKI_results1}, triangles represent individuals with the three largest HPD intervals. We observe that the anxiety score is underestimated for these people, but the high uncertainty informs us that we may need to pay additional attention to these individuals with severe anxiety issues; they might have been overlooked without uncertainty quantification.

\section{Discussion}

We propose a flexible approach, which can utilize image data as predictor variables via feature extraction through BayesCNN. The proposed method can simultaneously utilize predictors with different data structures, such as vector covariates and image data. Our study shows that the proposed method results in prediction performance comparable to that of existing deep learning algorithms in terms of prediction accuracy while enabling uncertainty quantification for estimation and prediction. By constructing an ensemble posterior through a mixture of feature posteriors, we can account for the uncertainties in feature extraction.

Our work is motivated by recently developed feature extraction approaches from DNN. These features can improve the performance of existing statistical approaches by accounting for complex dependencies among observations. For instance, \citet{bhatnagar2020computer} extracted features from the long-short-term memory (LSTM) networks and used them to capture complex inverse relationships in calibration problems. Similar to our approach, \citet{tran2020bayesian} used the extracted features from DNN as additional covariates in generalized linear mixed effect models. \citet{fong2021forward} proposed a nonlinear dimension reduction method based on deep autoencoders, which can extract interpretable features from high-dimensional data. 

The proposed framework can be extended to a wider range of statistical models, such as Cox regression in survival analysis or regression models in causal inference. Extracting features from other types of DNN is also available; for example, one can use CNN-LSTM \citep{wang2016dimensional} to capture spatio-temporal dependencies. Our ensemble approaches can significantly improve prediction and provide interpretable parameter estimates. 

\section*{Acknowledgement}
This work was partially supported by the National Research Foundation of Korea (2019R1A2C1007399, 2020R1C1C1A0100386814, 2022R1C1C1006735, RS-2023-00217705), ICAN (ICT Challenge and Advanced Network of HRD) support program (RS-2023-00259934) supervised by the IITP (Institute for Information \& Communications Technology Planning \& Evaluation). The authors are grateful to anonymous reviewers for their careful reading and valuable comments.

\section*{Data Availability}
The brain tumor images dataset is obtained from Kaggle (\url{https://www.kaggle.com/datasets/jakeshbohaju/brain-tumor/data}). The fMRI dataset for anxiety score and the malaria incidence datasets can be downloaded from \url{https://github.com/jeon9677/BayesCGLM}.

\bibliography{Reference}

\label{lastpage}

\section*{Supplementary Materials}

Web Appendices, Tables, and Figures referenced in Sections~\ref{BayesDeep}-\ref{applications}, as well as a zip file containing data and code, are available with this paper at the Biometrics website on Oxford Academic.

\vspace*{-8pt}

\end{document}


\maketitle

\section{Mathematical Details}
\subsection{Dropout as a Bayesian Approximation}
In this section, we describe results in \cite{gal2016dropout} that a deep neural network with dropout layers is mathematically equivalent to the approximation of the posteriors of the deep Gaussian process (GP). Specifically, we focus on the Gaussian response $\mathbf{y}$ case, but it can be easily extended to non-Gaussian responses. Consider deep neural networks (DNNs) with dropout layers described in the main manuscript. In a traditional approach, 

we can train the model by minimizing the following loss function with $L_2$ regularization terms:
\begin{equation}
     \mathcal{L}_{dropout} := -\frac{1}{2N}\sum_{n=1}^{N} ||\mathbf{y}_n - \mathbf{\widehat{y}}_n||_2^2+\sum_{l=1}^{L}\lambda^{\mathbf{w}}_{l}||\mathbf{W}_l||_2^2+\sum_{l=1}^{L}\lambda^{\mathbf{b}}_{l}||\mathbf{b}_l||_2^2.
\label{dropoutloss}
\end{equation}
Here $\lambda^{\mathbf{w}}_{l}$, $\lambda^{\mathbf{b}}_{l}$ are the shrinkage parameters for weight and bias parameters ($\mathbf{W}_l$ and  $\mathbf{b}_l$), respectively. Note that we can replace \eqref{dropoutloss} with other types of loss functions for non-Gaussian responses; for instance, we can use a sigmoid loss function for a binary classification problem. 

As described in the manuscript, the hierarchical neural network can be represented as a deep GP \citep{damianou2013deep}. Then, the Monte Carlo (MC) approximation of the log evidence lower bound (ELBO) of the deep GP \citet{gal2016dropout} is
\begin{equation}
\begin{split}
\mathcal{L}_{\text{GP-MC}} & :=\frac{1}{M}\sum_{m=1}^{M}\sum_{n=1}^{N} \log p(\mathbf{y}_n|\mathbf{x}_n,\lbrace \mathbf{W}_{l}^{(m)}, \mathbf{b}_{l}^{(m)} \rbrace_{l=1}^{L}) \\
&\quad -\text{KL}\Big(\prod_{l=1}^{L}q(\mathbf{W}_{l})q(\mathbf{b}_{l})\Big|\Big|\prod_{l=1}^{L}p(\mathbf{W}_{l})p(\mathbf{b}_{l})\Big),
\label{GPMCsuppl}
\end{split}
\end{equation} 
where $\lbrace\lbrace \mathbf{W}_{l}^{(m)}, \mathbf{b}_{l}^{(m)} \rbrace_{l=1}^{L}\rbrace_{m=1}^{M}$ is MC samples from the variational distribution $\prod_{l=1}^{L}q(\mathbf{W}_{l})q(\mathbf{b}_{l})$ defined in the manuscript. In \eqref{GPMCsuppl}, $p(\mathbf{W}_{l})$ and $ p(\mathbf{b}_l)$ denote independent standard normal priors for weight and bias parameters, respectively.

In Web Appendix D, we observe that even $M = 1$ can provide
reasonable approximations, though the results become more accurate with increasing $M$. 

As demonstrated in \cite{gal2016dropout}[Proposition 1], under the conditions of the $l$th layer having a large number of nodes $k_l$ and a small constant value $\sigma$ (a hyperparameter of the variational distribution controlling the spread of the distributions), $\text{KL}( q(\mathbf{W}_l)|| p(\mathbf{W}_l))$ can be approximated as

\begin{equation}\begin{split}
    \text{KL}( q(\mathbf{W}_l)|| p(\mathbf{W}_l)) &\approx \sum_{\forall i,j \in l} \frac{p_l}{2}((\mu^{w}_{l,ij})^{2}+(\sigma^{2} -(1+\log2\pi)-\log\sigma^{2})+C),
\end{split}
\label{KLdiversuppl1}
\end{equation}
where $\mu^{w}_{l,ij}$ is a variational parameter of the weight that controls the mean of the distribution, and $C$ is some constant. Similarly, $\text{KL}( q(\mathbf{b}_l)|| p(\mathbf{b}_l))$ can be approximated as
\begin{equation}\begin{split}
    \text{KL}( q(\mathbf{b}_l)|| p(\mathbf{b}_l)) &\approx \sum_{\forall i,j \in l} \frac{p_l}{2}((\mu^{b}_{l,ij})^{2}+(\sigma^{2} -(1+\log2\pi)-\log\sigma^{2})+C),
\end{split}
\label{KLdiversuppl2}
\end{equation}
where $\mu^{b}_{l,ij}$ is a variational parameter of the bias parameter that controls the mean of the distribution.
By plugging in these approximations to \eqref{GPMCsuppl}, we have
\begin{equation}  \begin{split}
     \mathcal{L}_{\text{GP-MC}} &\approx \sum_{n=1}^{N} \log N(y_n; \mathbf{W}_{n,L}\phi_{n,L-1}+\mathbf{b}_L,\tau^{-1}\mathbf{I}_N)\\ &-\sum_{l=1}^{L} \frac{p_l}{2}(||\bm{\mu}^{\mathbf{w}}_{l}||^{2}-k_l k_{l-1}(\sigma^{2} -(1+\log2\pi)-\log\sigma^{2})) \\ &-
     \sum_{l=1}^{L} \frac{p_l}{2}(||\bm{\mu}^{\mathbf{b}}_{l}||^{2}-k_l(\sigma^{2} -(1+\log2\pi)-\log\sigma^{2}))
\label{GPMCKLsuppl}
\end{split}
\end{equation}
with a constant precision $\tau$ (hyper parameter). 

By ignoring constant hyperparameter terms ($\tau$, and $\sigma$), the approximated version of the log evidence lower bound scaled by a positive constant $\frac{1}{\tau N}$ becomes  
\begin{equation} \begin{split}
     \mathcal{L}_{\text{GP-MC}} \approx -\frac{1}{2N}\sum_{n=1}^{N} ||\mathbf{y}_n - \mathbf{\widehat{y}}_n||_2^2 - \sum_{l=1}^{L}\frac{p_l}{2\tau N} ||\bm{\mu}^{\mathbf{w}}_{l}||_2^2 - \sum_{l=1}^{L}\frac{p_l}{2\tau N}||\bm{\mu}^{\mathbf{b}}_{l}||_2^2.
\label{GPMC1suppl}
\end{split}
\end{equation}
This implies that the posterior distribution of the weight parameter $w_{l,ij}$ is approximated with the mixtures of the spike distributions; one is centered around $\mu^{w}_{l,ij}$ and the other is centered around 0. Similarly, the posterior of the bias parameter $b_{l}$ is approximated through the spike distribution centered around $\mu^{b}_{l,i}$. The loss function \eqref{GPMC1suppl} becomes equivalent to \eqref{dropoutloss} by setting $\lambda^{\mathbf{w}}_l$ and $\lambda^{\mathbf{b}}_l$ as $\frac{p_l}{2\tau N}$. This implies that the frequentist NN with dropout layers is mathematically equivalent to the approximation of the posteriors of deep GP. Given this result, we can quantify uncertainty in the Bayesian network without requiring additional costs compared to the frequentist network. 

\subsection{Convolution Operation as an Affine Transformation}
\begin{figure}
  \begin{center}
      \includegraphics[width = 0.8 \textwidth]{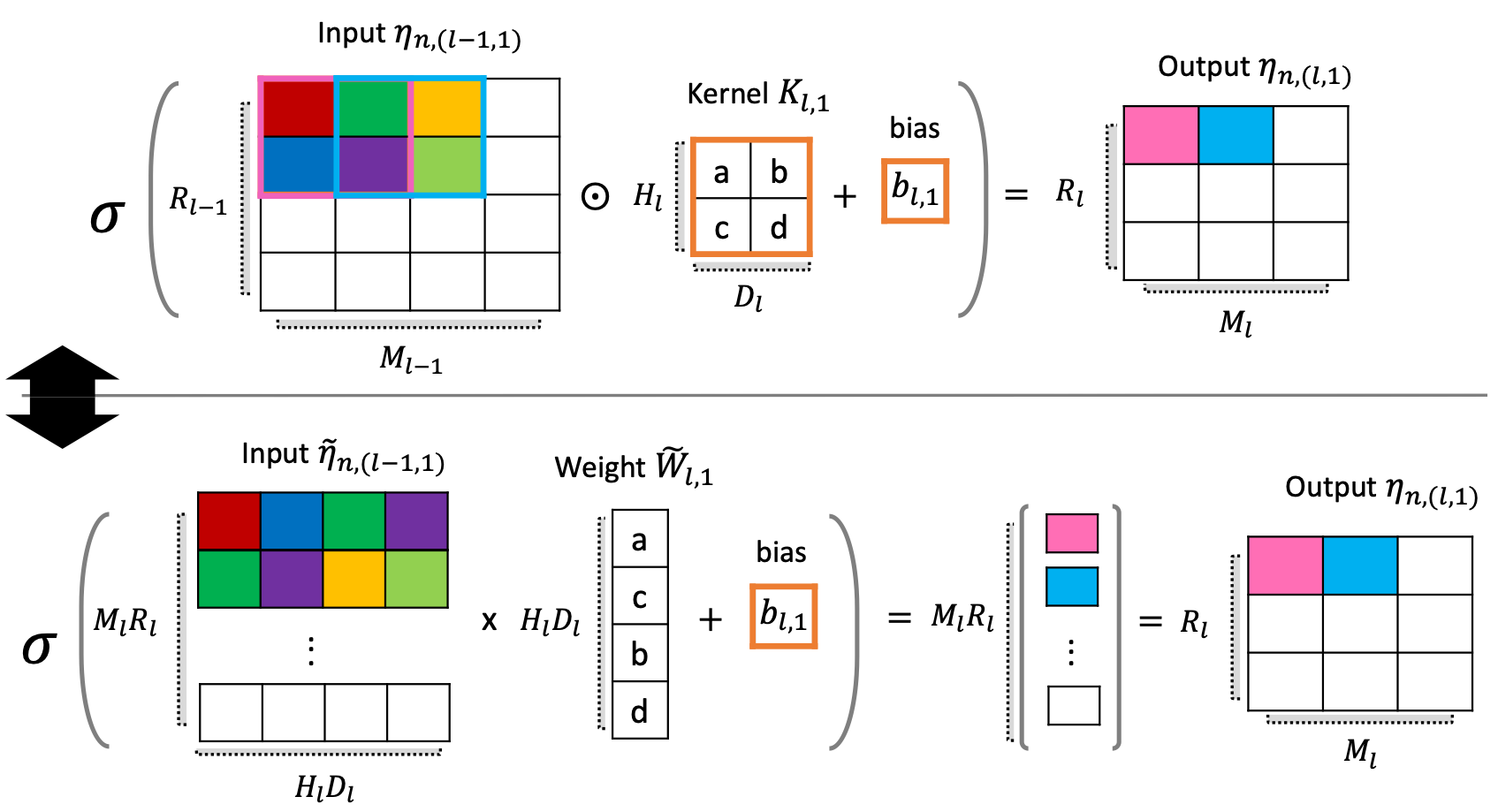}
  \end{center}
\caption[]{A graphical illustration for the convolution operation. The top panel illustrates the standard convolution operation. The bottom panel shows how the standard convolution can be reformulated as an affine transformation. Both are mathematically equivalent.}
\label{CNNmultiplication}
\end{figure}

Consider the same convolutional neural network (CNN) structure defined in the main manuscript. Without loss of generality, consider we have a single feature matrix $\bm{\eta}_{n,(l,1)}\in \mbR^{M_{l} \times R_{l}}$ with a kernel $\mathbf{K}_{l,1} \in \mbR^{H_l \times D_l}$ (i.e., $c=1$). As described in
Figure~\ref{CNNmultiplication}, we can rearrange the $(l-1)$th  input $\bm\eta_{n,(l-1,1)}$ as $\widetilde{\bm\eta}_{n,(l-1,1)} \in \mbR^{ M_lR_l \times H_{l}D_{l}}$ by vectorizing the $H_{l} \times D_{l}$ dimensional features from $\bm\eta_{n,(l-1,1)}$. Similarly, we can vectorize the kernel $\mathbf{K}_{l,1}$ and represent it as a weight matrix $\widetilde{\mathbf{W}}_{l,1} \in \mbR^{ H_{l} D_{l}}$. Then, we can perform matrix multiplication $\widetilde{\bm\eta}_{n,(l-1,1)}\widetilde{\mathbf{W}}_{l,1}\in \mbR^{M_lR_l}$ and rearrange it to obtain a matrix $\bm\eta_{n,(l,1)} \in \mbR^{M_{l} \times R_{l}}$, which is equivalent to output from the $l$th convolution layer. The above procedure can be generalized to the tensor input $\bm\eta_{n,l-1}=\lbrace \bm\eta_{n,(l-1,c)} \rbrace_{c=1}^{C_{l-1}} \in \mbR^{M_{l-1} \times R_{l-1} \times C_{l-1}}$ with $C_l$ number of kernels $\mathbf{K}_{l,c} \in \mbR^{H_{l} \times D_{l} \times C_{l-1}}$; therefore, the convolution operation is mathematically equivalent to standard DNN. 

\clearpage
\section{Sensitivity Analysis for Different Dropout Rates}

We conduct additional experiments to study the performance of BayesCGLM with varying dropout rates. For each real data application, we apply the three different dropout rates $\psi$ (0.15, 0.25, and 0.35). The remaining settings are identical to those given in the main manuscript. Tables~\ref{dropout_malaria}, \ref{dropout_braintumor}, and \ref{dropout_NKI} indicate that the results are similar across the different choices of dropout rates; we observe that the estimated regression coefficients and prediction performance are similar for all applications. In the main manuscript, we illustrate the model fitting results with a dropout rate of 0.25.

\begin{table}[tt]
\centering
\caption{Inference results for the malaria dataset with different dropout rates $\psi$. For all cases, the posterior mean of $\bm{\gamma}$, 95\% HPD interval, RMSPE, prediction coverage, and computing time (min) are reported in the table.}
\label{dropout_malaria}
\tabcolsep=0.03cm
\begin{tabular}{l l c c c}
& &$\psi=0.15$ &$\psi=0.25$ & $\psi=0.35$ \\ \hline
$\gamma_1$ (vegetation index) & Mean & 0.096  &  0.099 & 0.097  \\
& 95\% interval &  $(0.091, 0.101)$ & $(0.092, 0.107)$ &  (0.091 0.103)\\ 
$\gamma_2$ (proximity to water) & Mean & 0.077 & 0.074& 0.083 \\
& 95\% interval & $(0.072, 0.081)$ & $(0.068, 0.080)$  & $(0.078, 0.088)$  \\
$\gamma_3$ (rainfall) & Mean & 0.029 & 0.036 & 0.024 \\
& 95\% interval &  $(0.023, 0.036)$ & $(0.027,  0.045)$  & $(0.016,  0.03)$ \\
Prediction & RMSPE & 24.330   &  27.438 &  24.805  \\
& Coverage & 0.979  & 0.950 & 0.981 \\
Time (min)& &  57.421 & 57.518 & 56.119 \\
\hline
\end{tabular}
\end{table}

\begin{table}[tt]
\centering
\caption[]{Inference results for the brain tumor dataset with different dropout rates $\psi$. For all cases, the posterior mean of $\bm{\gamma}$, 95\% HPD interval, accuracy, recall, precision, and computing time (min) are reported in the table.}
\label{dropout_braintumor}
\tabcolsep=0.03cm
\begin{tabular}{l l c c c}
& &$\psi=0.15$ &$\psi=0.25$ & $\psi=0.35$ \\ \hline
$\gamma_1$ (first-order feature) & Mean & $-4.489$  & $-5.332$ & $-6.552$   \\
& 95\% interval & $(-5.536 , -3.660)$ & $(-7.049, -3.704)$&  $(-7.510, -5.526)$\\  
$\gamma_2$ (second-order feature) & Mean & 4.185  & 4.894 & 5.941 \\
& 95\% interval & $(3.318, 5.114)$ & $(3.303, 6.564)$  & $(4.851, 6.816)$ \\
Prediction & Accuracy & 0.904  & 0.924  & 0.934  \\
& Recall & 0.870 & 0.929 & 0.914\\
& Precision & 0.915 & 0.901 & 0.939 \\
Time (min) & &  294.457 & 293.533 & 293.192 \\
\hline
\end{tabular}
\end{table}

\begin{table}[tt]
\centering
\caption[]{The inference results for the brain fMRI dataset with different dropout rates $\psi$. For all cases, the posterior mean of $\bm{\gamma}$, 95\% HPD interval, RMSPE, prediction coverage, and computing time (min) are reported in the table.}
\label{dropout_NKI}
\tabcolsep=0.03cm
\begin{tabular}{l l c c c}
& &$\psi=0.15$ &$\psi=0.25$ & $\psi=0.35$ \\ \hline
$\gamma_1$ (neuroticism) & Mean &  2.591  &  3.496 & 2.423   \\
& 95\% interval &  $(1.587, 3.558)$  & $(2.629, 4.290)$ & $(1.438, 3.277)$ \\ 
$\gamma_2$ (extraversion)& Mean &  $-0.997$ & $-1.123$ & $-1.326$ \\
& 95\% interval & $(-1.748,-0.188)$ & $(-1.829,-0.386)$  & $(-2.061, -0.722)$  \\
$\gamma_3$(agreeableness) & Mean & 0.617 & 1.113 &  0.243 \\
& 95\% interval & $(-0.115, 1.369)$  & $(0.480, 1.760)$  & $(-0.415, 0.848)$ \\
$\gamma_4$(conscientiousness) & Mean & $-1.025$ & $-1.394$ & $-1.340$\\
& 95\% interval & $(-1.902, -0.106)$  & $(-2.226,-0.507)$  & $(-2.158, -0.485)$  \\
Prediction & RMSPE &  8.955  &  9.030 & 9.347 \\
& Coverage & 0.942  & 0.923 & 0.923 \\
Time (min) & &  3.148 & 3.459 & 3.311 \\
\hline
\end{tabular}
\end{table}

\section{Implementation Details}
\subsection{BayesCNN Implementation}
\begin{figure}[htbp]
\begin{center}
\includegraphics[width = 1\textwidth]{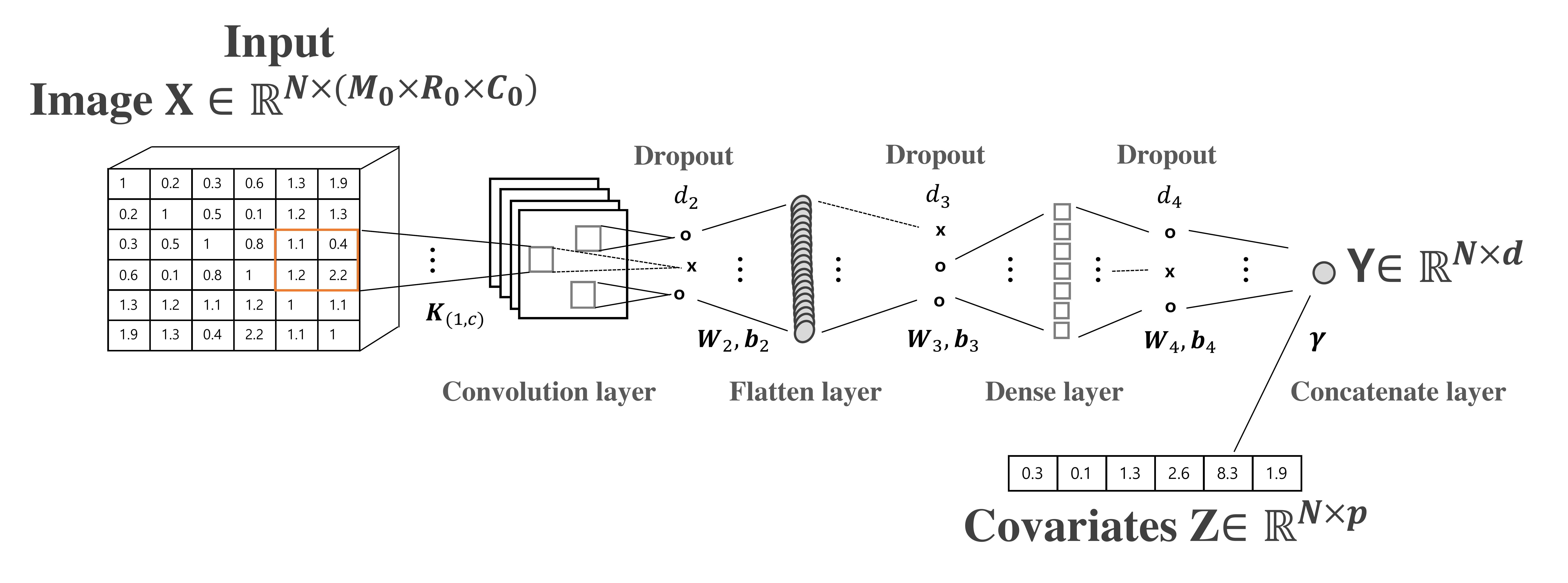}
\end{center}
\caption[]{Illustration for the BayesCNN with four layers.}
\label{bayesCNN_fig}
\end{figure}

In this section, we describe implementation details for BayesCNN \citep{gal2016bayesian}. For example, consider BayesCNN with four layers: a convolution layer, a flatten layer, a dense layer, and a concatenate layer. Then $\bm\theta= \lbrace (\mathbf{W}_l \in \mbR^{k_l \times k_{l-1}}, \mathbf{b}_l \in \mbR^{k_l}, \bm{\gamma} \in \mbR^p),l=1,2,3,4 \rbrace$ is a set of parameters. Note that we define $\bm{\gamma}$ separately from $\mathbf{W}_l$ to represent coefficients corresponding to the vector covariates $\mathbf{Z}$. We present a graphical description of BayesCNN in Figure~\ref{bayesCNN_fig}.

Following the notation in Section 3.1, we can vectorize the input kernel $\mathbf{K}_{(1,c)}$, where $c=1,\cdots, C_0$ as a weight matrix $\mathbf{W}_1 \in \mbR^{H_1 D_1 C_0}$. The output from the convolution layer $\bm{\eta}_{n,(1,c)} \in \mbR^{M_1 \times R_1 \times C_1}$ can also be vectorized as $\bm\phi_{n,1} \in \mbR^{M_1 R_1 C_1}$, where $k_0 = H_1 D_1 C_0$ and $k_1 = M_1 R_1 C_1$. In the training step, we apply dropout $\mathbf{d}_l \in \mbR^{k_l}, l=2,3,4$ to the features, the output from each previous layer. The dropout randomly removes the connections from the nodes of the features with a pre-specified dropout rate $\psi$ for each iteration. This step is repeated until the parameters $\bm{\theta}$ converge. Algorithm~\ref{algo_bayesCNN2} provides detailed steps of BayesCNN.

\begin{algorithm}
\caption{BayesCNN algorithm \citep{gal2016bayesian}}\label{algo_bayesCNN2}
\textbf{Part1. Training} \\
\small \textbf{Input}: $\mathbf{X} \in \mbR^{N\times (M_0 \times R_0 \times C_0)}$ (image or correlated input), $\mathbf{Y} \in \mbR^{N\times d}$ (response), $\mathbf{Z} \in \mbR^{N \times p} $ (scalar covariates), $\psi$ (dropout rate) \\
\textbf{Output}: Estimated parameters $\widehat{\bm\theta}= \lbrace (\mathbf{\widehat{W}}_l, \mathbf{\widehat{b}}_l, \bm{\widehat{\gamma}}),l= 1,2,3,4 \rbrace$  
\begin{algorithmic}[1]
    \Repeat
        \State \textbf{Forward Propagation:}
        \State Generate $\mathbf{d}_l$ from  Bernoulli($\psi$), $l=2,3,4$
         \For {$n=1,2,\ldots,N$}
         \State \small Set $[\bm{\eta}_{n,(1,c)}]_{m,r} = \sigma_1 \Big(\sum_{i=1}^{H_{1}}\sum_{j=1}^{D_{1}}\sum_{k=1}^{C_0} ([\mathbf{K}_{1,c}]_{i,j,k} [\mathbf{X}_{n}]_{m+i-1,r+j-1,k}) + b_{1,c}\Big)$
        \State Set $\sigma_1(\widetilde{\mathbf{X}}_n{\mathbf{W}}_1 + \mathbf{b}_1) = \lbrace \text{vec}(\bm{\eta}_{n,(1,c)}) \rbrace_{c=1}^{C_0}$
         \State Set $\bm\phi_{n,1} =  \sigma_1(\widetilde{\mathbf{X}}_n{\mathbf{W}}_1 + \mathbf{b}_1)$
         \State  Set $\bm\phi_{n,2} = \sigma_2 \Big( \mathbf{W}_{2}\bm\phi_{n,1}\circ\mathbf{d}_{2} + \mathbf{b}_{2} \Big)$
        \State  Set $\bm\phi_{n,3} = \sigma_3 \Big( \mathbf{W}_{3}\bm\phi_{n,2}\circ\mathbf{d}_{3} + \mathbf{b}_{3} \Big)$
         \State Set $\mathbf{o}_{n} =\sigma_4 \Big(\mathbf{W}_{4}\bm\phi_{n,3}\circ \mathbf{d}_4 + \mathbf{b}_{4} \Big) + \bm{\gamma}\mathbf{z}_n$
    \EndFor 
   \State \textbf{Backward Propagation:}
\State Update $\bm\theta$ with ADAM optimizer 
    \Until $\bm\theta$ converges;
\end{algorithmic}
\normalsize
\hrulefill \\
\textbf{Part2. Prediction}\\
\small
\textbf{Input}: $\mathbf{X}^{\ast} \in \mbR^{N_{\text{test}} \times (M_0 \times R_0 \times C_0)}$ (image or correlated input), $\mathbf{Z}^{\ast} \in \mbR^{N_{\text{test}} \times p}$ (scalar covariates) \\
\textbf{Output}: Predictive distribution of 
$\widehat{\mathbf{o}}_{n}^{\ast}$ 
\begin{algorithmic}[1]
     \For {$m=1,2,\ldots,M$}
        \State Generate $\mathbf{d}_l$ from Bernoulli($\psi$), $l=2,3,4$
         \For {$n=1,2,\ldots,N_{\text{test}}$}
         \State Set \small $[\bm{\eta}^{\ast}_{n,(1,c)}]_{m,r} = \sigma_1 \Big(\sum_{i=1}^{H_{1}}\sum_{j=1}^{D_{1}}\sum_{k=1}^{C_0} ([\mathbf{\widehat{K}}_{1,c}]_{i,j,k} [\mathbf{X}^{\ast}_{n}]_{m+i-1,r+j-1,k}) + \widehat{b}_{1,c}\Big)$
          \State Set $\sigma_1(\widetilde{\mathbf{X}}_n^{\ast}\widehat{{\mathbf{W}}}_1 + \mathbf{b}_1) = \lbrace \text{vec}(\bm{\eta}^{\ast}_{n,(1,c)}) \rbrace_{c=1}^{C_0}$
         \State Set $\bm\phi^{\ast (m)}_{n,1} =\sigma_1(\widetilde{\mathbf{X}}_n^{\ast}\widehat{\mathbf{W}}_1 + \mathbf{b}_1)$
         \State Set $\bm\phi^{\ast (m)}_{n,2} = \sigma_2 \Big(\mathbf{\widehat{W}}_{2}\bm\phi^{\ast (m)}_{n,1}\circ\mathbf{d}_{2} + \mathbf{\widehat{b}}_{2} \Big)$
        \State Set $\bm\phi^{\ast (m)}_{n,3} = \sigma_3 \Big(\mathbf{\widehat{W}}_{3}\bm\phi^{\ast (m)}_{n,2}\circ\mathbf{d}_{3} + \mathbf{\widehat{b}}_{3} \Big)$
         \State Set $\widehat{\mathbf{o}}_{n}^{\ast(m)} = \sigma_4 \Big(\mathbf{\widehat{W}}_{4}\bm\phi^{\ast (m)}_{n,3}\circ \mathbf{d}_4 + \mathbf{\widehat{b}}_{4} \Big) + \bm{\widehat{\gamma}}\mathbf{z}_n$
    \EndFor 
\EndFor 
 \State Construct the predictive distribution from $M$ number of MC samples: $\lbrace \widehat{\mathbf{o}}_{n}^{\ast (m)} \rbrace_{m=1}^{M}$
\end{algorithmic}
\end{algorithm}

After we obtain estimated parameters $\widehat{\bm\theta}$ from the training step (Part 1 in Algorithm~\ref{algo_bayesCNN2}), BayesCNN provides predictive distribution of the outputs from MC sampling (Part 2 in Algorithm~\ref{algo_bayesCNN2}). For a given image $\mathbf{X}^{\ast}$ and vector covariates $\mathbf{Z}^{\ast}$, the predictive distribution of output is obtained via MC dropout. For each MC iteration, $\lbrace \mathbf{d}_l \rbrace_{l=2}^{L}$ are randomly sampled from the Bernoulli distribution with rate $\psi$ and are applied to the features; they are not the same as the dropout vectors used during the training step. We can repeat this $M$ times to obtain an MC sample of $M$ predicted outputs.

\subsection{BayesCGLM Algorithm}
\begin{figure}                      
    \begin{center}
\includegraphics[width = 0.9\textwidth]{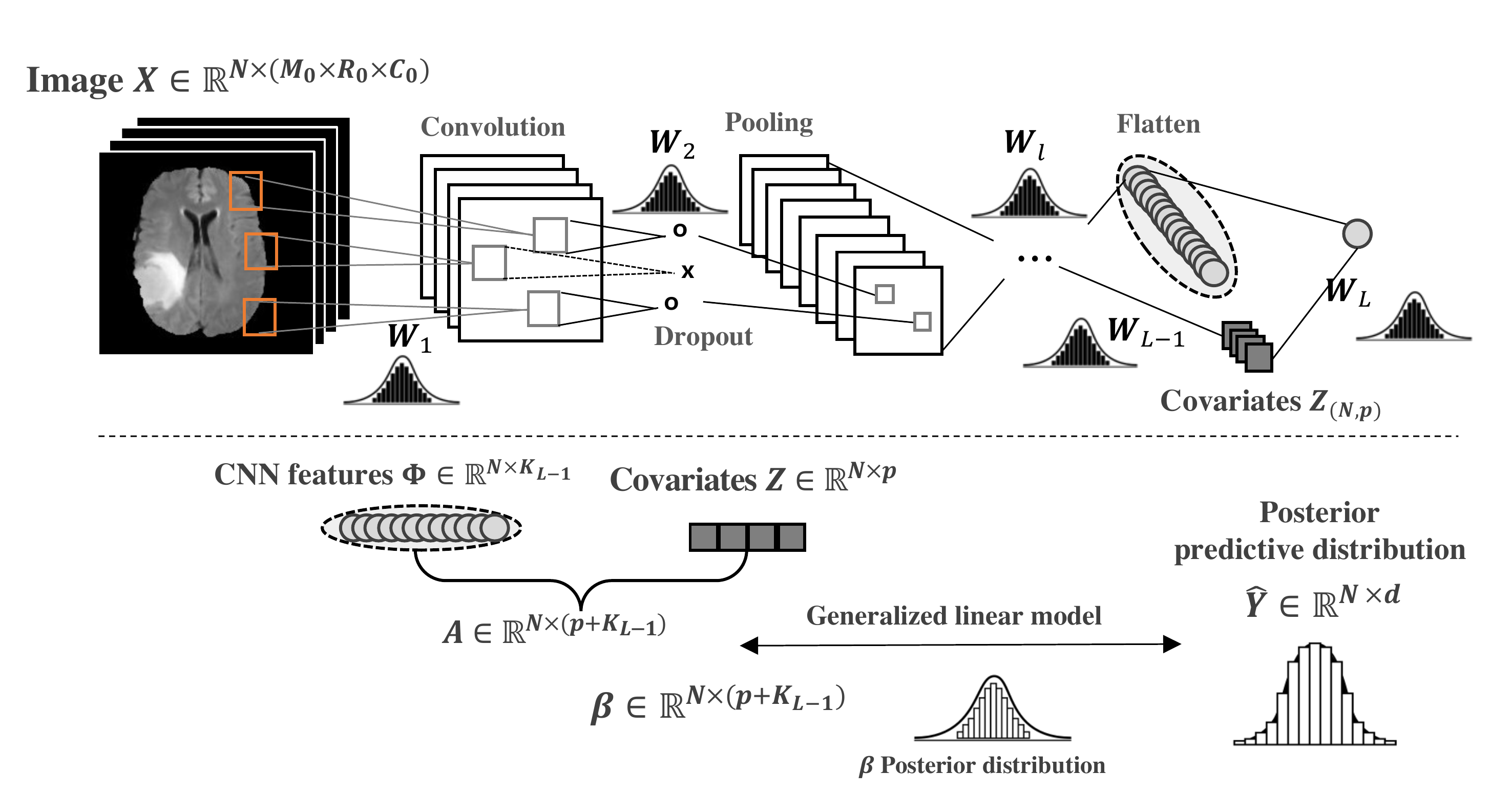}
\end{center}
\caption[]{A graphical illustration for BayesCGLM. The top panel shows how CNN is used to extract relevant features using the MC dropout. The bottom panel shows that the extracted features and the additional covariates $\mathbf{Z}$ are used as predictors for GLM, constructing the posterior distribution for the coefficients and the posterior predictive distribution.}
\label{BayesCGLM}
\end{figure}

In this section, we describe the algorithm of BayesCGLM (Algorithm~\ref{algo}). A graphical description of \noindent BayesCGLM is presented in Figure~\ref{BayesCGLM}. 

\begin{algorithm}
	\caption{BayesCGLM algorithm}
\textbf{Input}: $\mathbf{X}$ (image or correlated input), $\mathbf{Y}$ (response), $\mathbf{Z}$ (scalar covariates)

\textbf{Output}: Posterior distribution of model parameters.
	\begin{algorithmic}[1]
	\State Fit BayesCNN by using $\mathbf{X}$ and $\mathbf{Z}$ as input and $\mathbf{Y}$ as output. Then, for each $m$
		\For {$m=1,2,\ldots,M$}
			\State Extract features $\bm{\Phi}^{(m)}$ from the last hidden layer.
			\State Fit GLM by regressing $\mathbf{Y}$ on $[\mathbf{Z}, \bm{\Phi}^{(m)}]$ through the Laplace approximation 
			\State Obtain the $m$th feature-posterior distribution of the model parameter.
		\EndFor
	\State Construct an ensemble posterior by aggregating $M$ number of feature-posteriors.
	\end{algorithmic} 
	\label{algo}
\end{algorithm}

\newpage
\section{Simulation Studies}\label{Sec:Simul}

In this section, we apply BayesCGLM to three different simulated examples, with a Gaussian response (described in the \ref{Gaussian_case}), binary response (described in the \ref{binary_case}), and Poisson response (described in the \ref{Poisson_case}). We implement our approach in {\tt{TensorFlow}}, an open-source platform for machine learning. Parallel computation is implemented through the {\tt{multiprocessing}} library in {\tt{Python}} (https://docs.python.org/3/
library/multiprocessing.html). The computation times are based on 8-core AMD Radeon Pro 5500 XT processors. The configuration of the CNN structures, including the loss function, activation function, and tuning details used in our experiments, are provided in the \ref{Sec:CNNstrucSimu} of the Supplementary Materials.

\subsection{Simulation Experiment}
We generate 1,000 lattice datasets as images, which have a spatially correlated structure given by the Gaussian process. The responses are created by the canonical link function $g^{-1}$ and the natural parameter $\bm{\lambda}=g^{-1}(\bm\Phi + \mathbf{Z}\bm{\gamma})$ with $\bm{\gamma}=(1,1)$. Here, $\bm\Phi$ indicates the vectorized local features of the images, $\mathbf{Z}$ represents covariates generated from a standard normal distribution, and $\bm{\gamma}$ are the true coefficients. 

\begin{figure}
    \begin{center}
  \includegraphics[width = 0.9\textwidth]{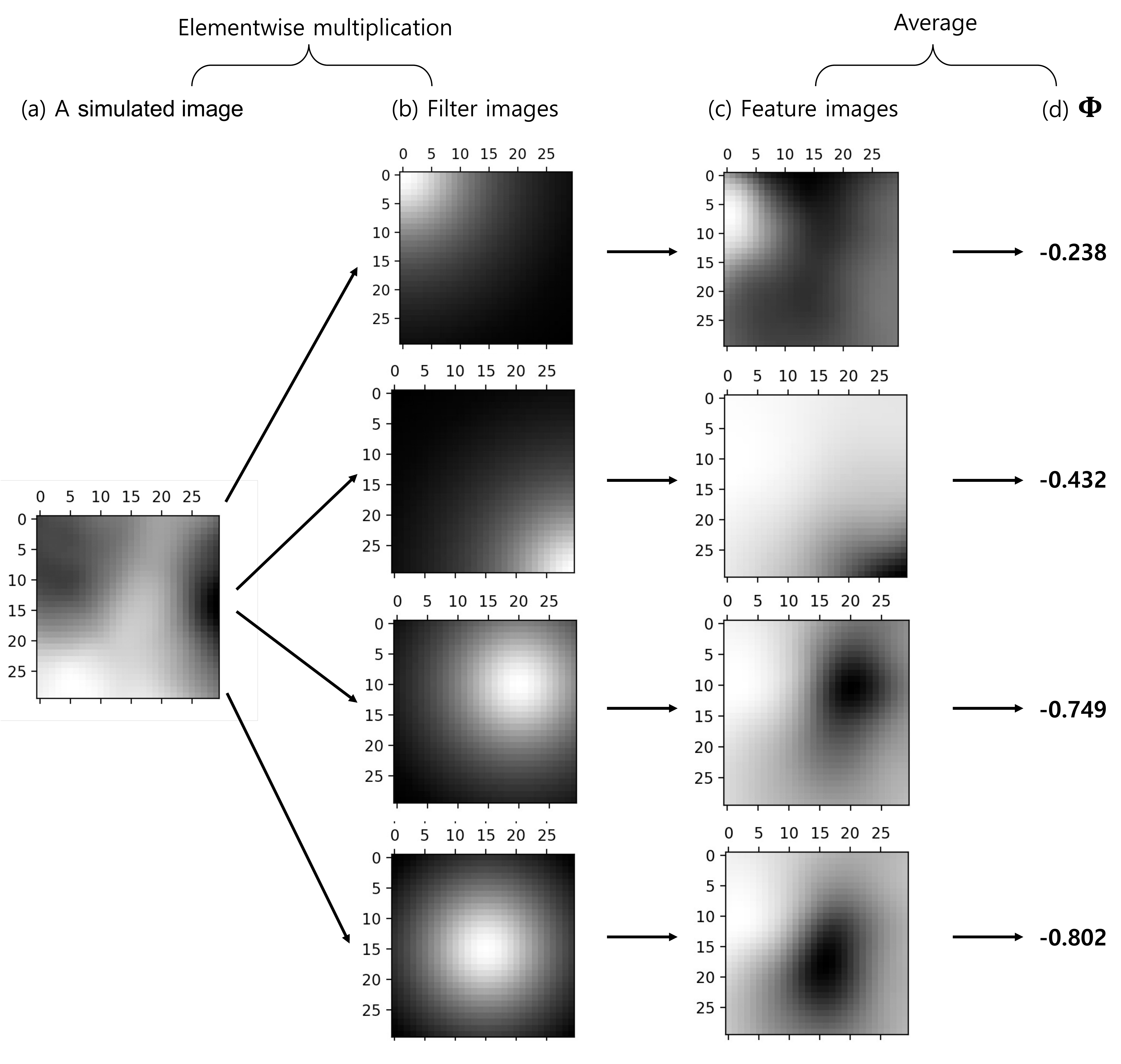}
    \end{center}
\caption{(a) An example of a generated $30 \times 30$ size image, $\mathbf{X}_n$. (b) Images of four filters were used in our simulation studies. Each of the filters highlights top-left ($\mathbf{K}_1$), bottom-right ($\mathbf{K}_2$), top-right ($\mathbf{K}_3$), and center ($\mathbf{K}_4$) respectively. (c) Feature images are generated by implementing elementwise multiplication between the generated sample image and the filter images. For all figures, the darker gray color corresponds to the smaller pixel values. (D) Each number is the resulting feature $\bm{\phi}$, calculated by averaging over all the pixels within the feature image.}
\label{Simulated_images}
\end{figure}

\paragraph{Simulation Design}
We first set up spatial locations  $\mathbf{s}_1,\cdots,\mathbf{s}_{900}$ on a $30\times 30$ regular lattice with a spacing of one. We generate spatially correlated data from a Gaussian process with mean 0 and the Mat\'{e}rn class covariance function \citep{stein2012interpolation} 
\[ \varphi(d) = \frac{\sigma^2}{2^{\nu-1}\Gamma(\nu)} (d/\rho)^{\nu}K_\nu(d/\rho),
\] 
where $d$ is the Euclidean distance between two points. Here $\sigma^2$, $\nu$, and $\rho$ denote the variance, smoothness, and range parameters, respectively. In our simulation studies, we set $\sigma^2=1,\nu=0.5$ and $\rho=15$. We repeat the above procedure for 1,000 times to generate $1,000$ number of image observations $\mathbf{X} = \lbrace\mathbf{X}_n \rbrace_{n=1}^{1000}$. Figure~\ref{Simulated_images} (a) illustrates a single realization of the simulated image. 

Then we extract features from simulated images through a $C$ number of different filters $\mathbf{K}_c\in \mbR^{30 \times 30}$ for $c=1,\cdots,C$ (Figure \ref{Simulated_images} (b)). In our study, we used $C=4$ for Gaussian and binary cases. For the Poisson case, we use only two filters ($\mathbf{K}_1$ and $\mathbf{K}_2$, which extract features in the top-left area and the bottom-right area, respectively) to extract features. This is because using all four filters leads to a large dispersion of Poisson responses. The weights in each filter are designed to capture local characteristics in different parts of the input images. Specifically, we use inverse quadratic basis functions $\varphi(d) = 1/(1+(\delta d)^2)$, where $d$ is the Euclidean distance between pixels in $\mathbf{K}_c$ and a focal point $\mathbf{u}_c$, and $\delta$ is the tuning parameter that controls the decay of the kernel as the distance grows. Here, we set $\mathbf{u}_1=(0,0)$, $\mathbf{u}_2=(10,20)$, $\mathbf{u}_3=(15,15)$, $\mathbf{u}_4=(30,30)$ to extract local features in different areas, and $\delta=0.1$ for all $c=1,\dots,4$. Based on the filters, we can extract features $\Phi_{n,c}$ as
\[
\Phi_{n,c} = \frac{1}{30}\frac{1}{30}\sum_{i=1}^{30} \sum_{j=1}^{30} \mathbf{K}_{c,ij} \mathbf{X}_{n,ij},
\]
for $c=1,\cdots,C$. Each $\mathbf{K}_c$ extracts a local feature of the images in different areas, such as bottom-left, bottom-right, center, or top-right. Therefore, the whole feature vector is $\bm\Phi_n=\lbrace \Phi_{n,c} \rbrace_{c=1}^{C} \in \mbR^{C}$ (Figure \ref{Simulated_images} (d)).

Finally, we generate covariates $\mathbf{Z} =\lbrace \mathbf{z}_{n} \rbrace_{n=1}^{1000} \in \mbR^{1000\times 2}$, where individual elements are sampled from a standard normal distribution. For given the generated features $\bm\Phi=\lbrace \bm\Phi_{n} \rbrace_{n=1}^{1000} \in \mbR^{1000\times C}$ above, we calculate $\bm{\lambda}=g^{-1}(\bm\Phi + \mathbf{Z}\bm{\gamma})$ with $\bm{\gamma}=(1,1)$, i.e., the true values of the regression coefficients for the covariates are 1's. We simulate $\mathbf{Y} \sim N(\bm{\lambda},\mathbf{I})$ for normal, $\mathbf{Y} \sim \text{Poisson}(\bm{\lambda})$ for count, and $\mathbf{Y} \sim \text{Bernoulli}(\bm{\lambda})$ for binary cases. Note that the features $\bm\Phi_{n}$'s used to generate $\mathbf{Y}$ are not used for model fitting. Our BayesCGLM uses the generated images $\mathbf{X}_n$ and $\mathbf{z}_n$ as input variables and conducts feature extraction on $\mathbf{X}_n$ by itself.  We use the first $N=700$ observations for model fitting and the remaining $N_{\text{test}}=300$ for performance testing.
To measure prediction accuracy, we calculate the root mean square prediction error (RMSPE) and the empirical coverage of prediction intervals. We also report the average computing time from simulations. For each simulation setting, we repeat the simulation 500 times.

\paragraph{Comparative Analysis}
To demonstrate the performance of our approach, we compare the parameter estimation and response prediction performance of the proposed BayesCGLM with Bayesian CNN (BayesCNN) \citep{gal2016bayesian} and generalized linear model (GLM). In the training step, we use the standard early stopping rule based on prediction accuracy to avoid overfitting issues. We train CNN by using both images $\mathbf{X}$ and covariates $\mathbf{Z}$ as input and response $\mathbf{Y}$ as output. By placing $\mathbf{Z}$ at the last fully connected layer, we can obtain the weight parameters that correspond to the regression coefficients $\bm{\gamma}$ in $g(E[\mathbf{Y}|\mathbf{Z},\bm{\Phi}^{(m)}]) = \mathbf{Z}\bm{\gamma}_m + \bm{\Phi}^{(m)}\bm{\delta}_m$; note that this can only provide point estimates of the weights. Since it is challenging to use images as predictors in the GLM, we fit the GLM by regressing $\mathbf{Y}$ on $\mathbf{Z}$ without using images.

\subsection{Gaussian Case}\label{Gaussian_case}
We first describe the simulation study results when the response $\mathbf{Y}$ is generated from a Gaussian distribution, i.e., $\mathbf{Y} \sim N_{1000}(\bm{\lambda},\mathbf{I})$. For MC dropout sampling, we set $M=300$, which is large enough to represent the estimation and prediction uncertainties for all the simulated and real data examples presented in the main document without requiring too much computational time; increasing $M$ to a bigger number did not lead to improvement in the prediction accuracy and the uncertainty quantification performance. We also compare the results with a single dropout sample (i.e., $M=1$) to examine how much improvement in inference performance can be attributed to the repeated dropout sampling. 

\begin{figure}[htbp]
\begin{center}
\includegraphics[width = 1\textwidth]{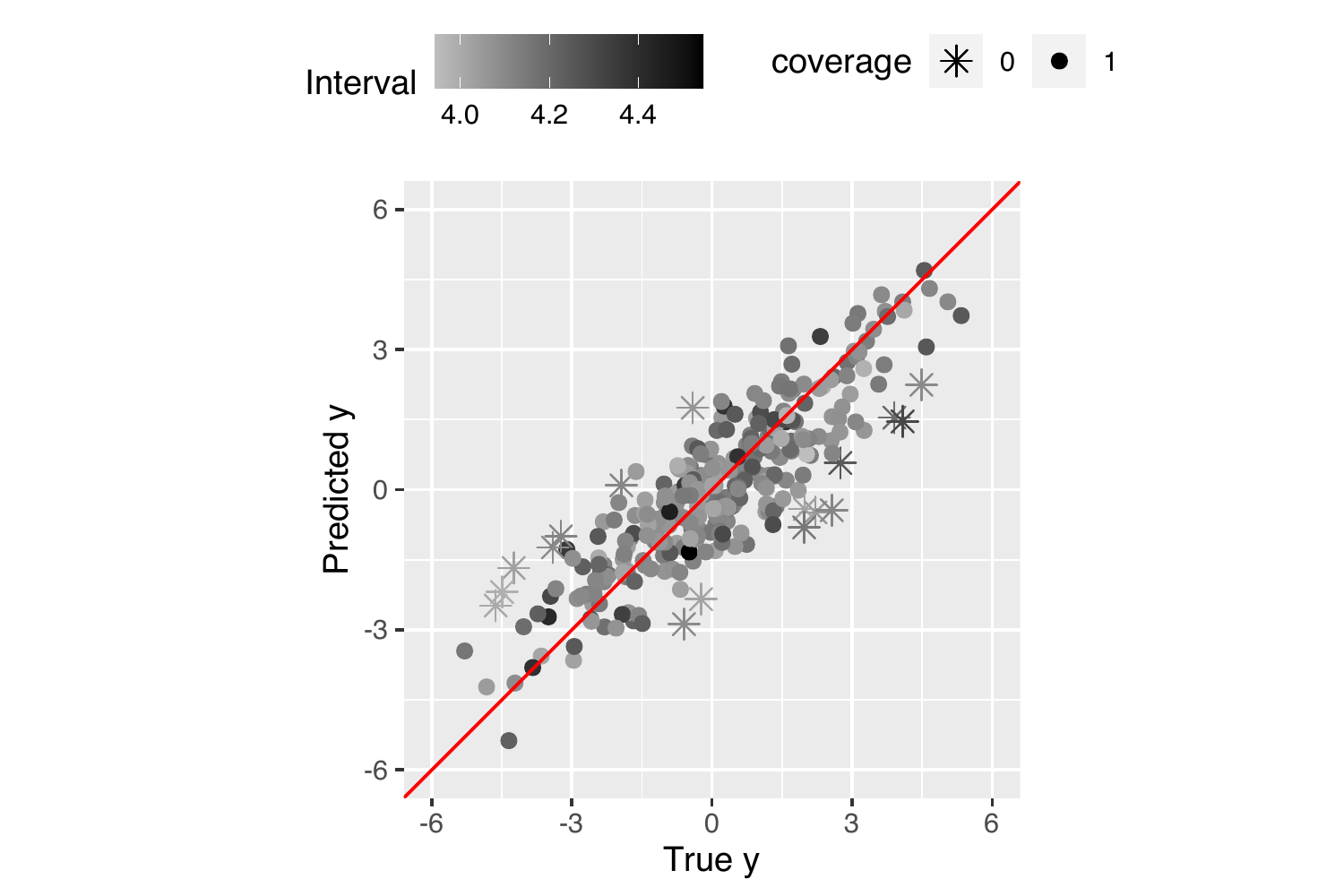}
\end{center}
\caption[]{Comparison between the true and predicted Gaussian responses from BayesCGLM. Circles (asterisks) represent that the true responses are (not) covered in 95\% HPD intervals.}
\label{simul_pred_normal}
\end{figure}

Figure \ref{simul_pred_normal} illustrates the agreement between the true and predicted responses in a simulated dataset. Since BayesCGLM can quantify uncertainties in predictions, we also visualize the 95\% highest posterior density (HPD) prediction intervals, which include the true $\mathbf{Y}$ well. 

Table \ref{simul_normal} reports the inference results from different methods. We observe that the parameter estimates obtained from BayesCGLM and GLM are similar to the simulated truth on average, while the estimates from BayesCNN \citep{gal2016bayesian} are biased. As we have pointed out in Section 3.2, simultaneous updates of high-dimensional weight parameters through a stochastic gradient descent algorithm can lead to inaccurate estimates of $\bm{\gamma}$. 

Our BayesCGLM with $M=300$ results in accurate uncertainty quantification, yielding credible intervals for the regression coefficients $\bm{\gamma}$ whose empirical coverage is closer to the nominal coverage (95\%) than the BayesCGLM with $M=1$ and GLM. This shows that both the repeated MC dropout sampling and the use of extracted features improve uncertainty quantification performance for $\bm{\gamma}$. 
Furthermore, the prediction accuracies (RMSPE, prediction coverage) from BayesCGLM are comparable to BayesCNN. On the other hand, the RMSPE obtained from GLM is much higher than those from the other deep learning approaches.

\begin{table}[tt]
\centering
\caption{Inference results for the simulated Gaussian datasets with the sample of size $500$. For all methods, posterior mean, RMSPE, and computing time (min) are averaged
from 300 repeated simulations. Estimation and prediction coverages are also reported.
The numbers in the parentheses indicate standard deviations obtained from the repeated
simulations.}
\label{simul_normal}
\begin{tabular}{l l c c c c}
\hline
 \multicolumn{2}{c}{} & \multicolumn{2}{c}{\textbf{BayesCGLM}} & \textbf{BayesCNN} & \textbf{GLM} \\ 
\cmidrule(lr){3-4} \cmidrule(lr){5-5}
 & &  $M=300$ & $M= 1$ & $M=300$ \\
\hline
$\gamma_1$ & Mean & 0.993 (0.041) &  0.954 (0.058)  & 0.888 (0.093) &  1.003 (0.057) \\
& Coverage & 0.944 & 0.918 & - & 0.930 \\  
$\gamma_2$ & Mean & 0.992 (0.041) & 0.946 (0.060) & 0.883 (0.104) & 1.000 (0.055)  \\
& Coverage & 0.954 & 0.924  &  - & 0.912 \\  
Prediction & RMSPE & 1.048 (0.045) & 1.044 (0.045)  & 1.076 (0.053)  & 1.416 (0.058)  \\       
& Coverage & 0.940  &  0.930  & 0.947   & 0.975 \\    
Time (min) & & 35.725  & 19.037  & 18.894   & 0.003 \\   \hline 
\end{tabular}
\end{table}

To investigate the performance of algorithms with a larger sample size, we repeat a simulation study with samples of size 5,000. Here, we use the same CNN structure and hyperparameters as before. Both BayesCGLM and BayesCNN work well in terms of estimation and prediction (Table~\ref{simul_normal_bigsample}). We observe that BayesCGLM with $M=300$ achieves a 95\% nominal rate for prediction coverage. Furthermore, BayesCGLM can provide uncertainty quantification of regression coefficients, while BayesCNN cannot.

\begin{table}[tt]
\centering
\caption{Inference results for the simulated Gaussian datasets with the sample of size $5,000$. For all methods, posterior mean, RMSPE, and computing time (min) are averaged from 300 repeated simulations. Estimation and prediction coverages are also reported. The numbers in the parentheses indicate standard deviations obtained from the repeated simulations.}
\label{simul_normal_bigsample}
\begin{tabular}{l l c c c c}
\hline
 \multicolumn{2}{c}{} & \multicolumn{2}{c}{\textbf{BayesCGLM}} & \textbf{BayesCNN} & \textbf{GLM} \\ 
\cmidrule(lr){3-4} \cmidrule(lr){5-5}
 & &  $M=300$ & $M= 1$ & $M=300$ \\
\hline
$\gamma_1$  & Mean & 0.999 (0.017) &  1.000 (0.19)  & 0.993 (0.020) & 1.002  (0.025) \\
& Coverage & 0.967 & 0.960 & - & 0.937 \\  
$\gamma_2$ & Mean & 0.998 (0.018) & 1.000 (0.19) & 0.991 (0.021) & 1.000 (0.025)  \\
& Coverage & 0.943  & 0.950 &  - & 0.937 \\ Prediction & RMSPE & 1.021 (0.017) & 1.08 (0.032)  & 1.024 (0.017)  & 1.418 (0.026)  \\       
& Coverage & 0.959 & 0.922  & 0.922  & 0.940 \\    
Time (min) & & 58.561 & 22.836   &  21.002 & 0.003 \\
\hline
\end{tabular}
\end{table}

\subsection{Binary Case} \label{binary_case}
\begin{figure}[htbp]
\centering\begin{subfigure}{.85\textwidth}\includegraphics[width=1\linewidth]{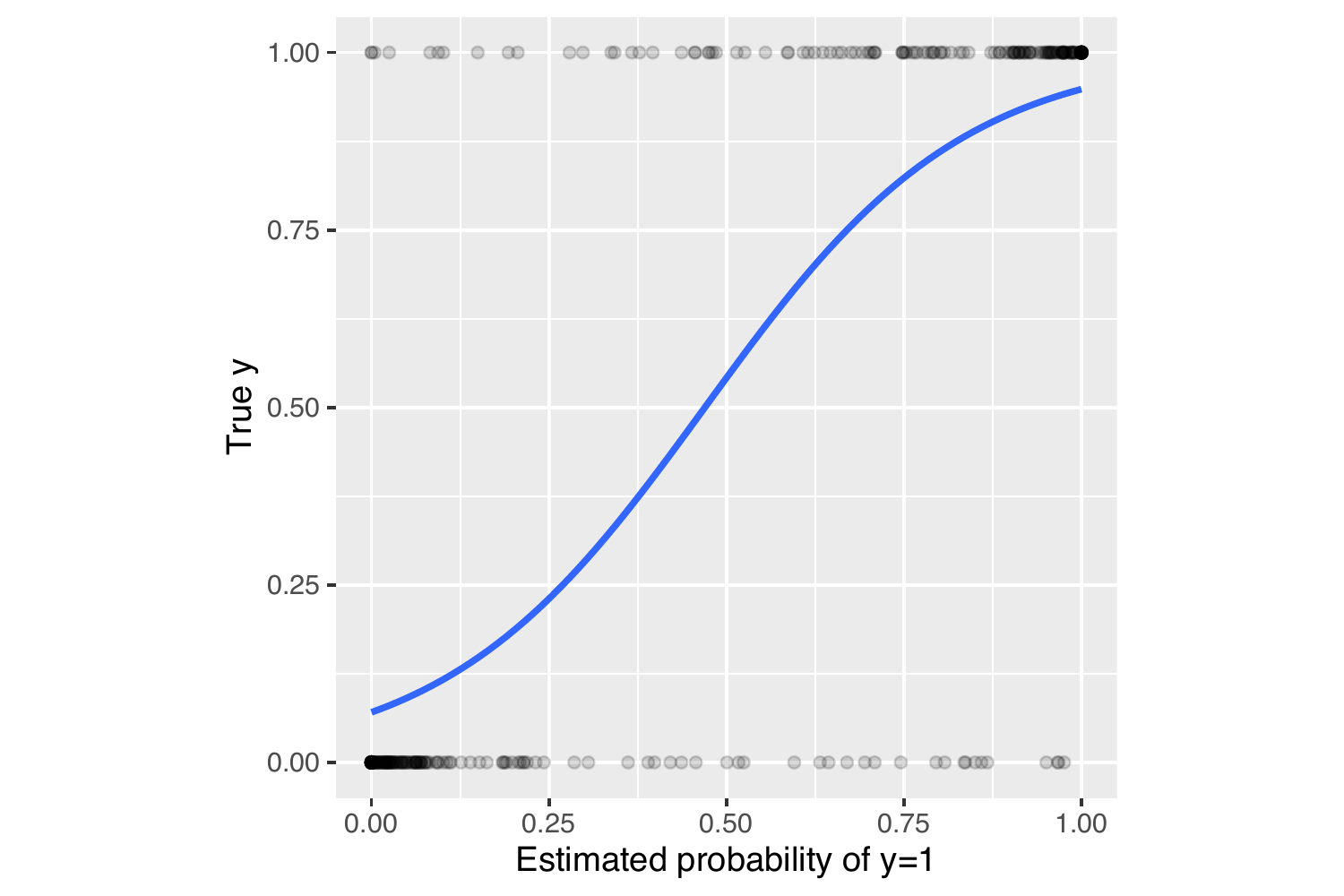}
  \caption{A logistic regression curve}
  \label{simul_pred_binary1}
\end{subfigure}%
\\
\centering\begin{subfigure}{.95\textwidth}\includegraphics[width=0.98\linewidth]{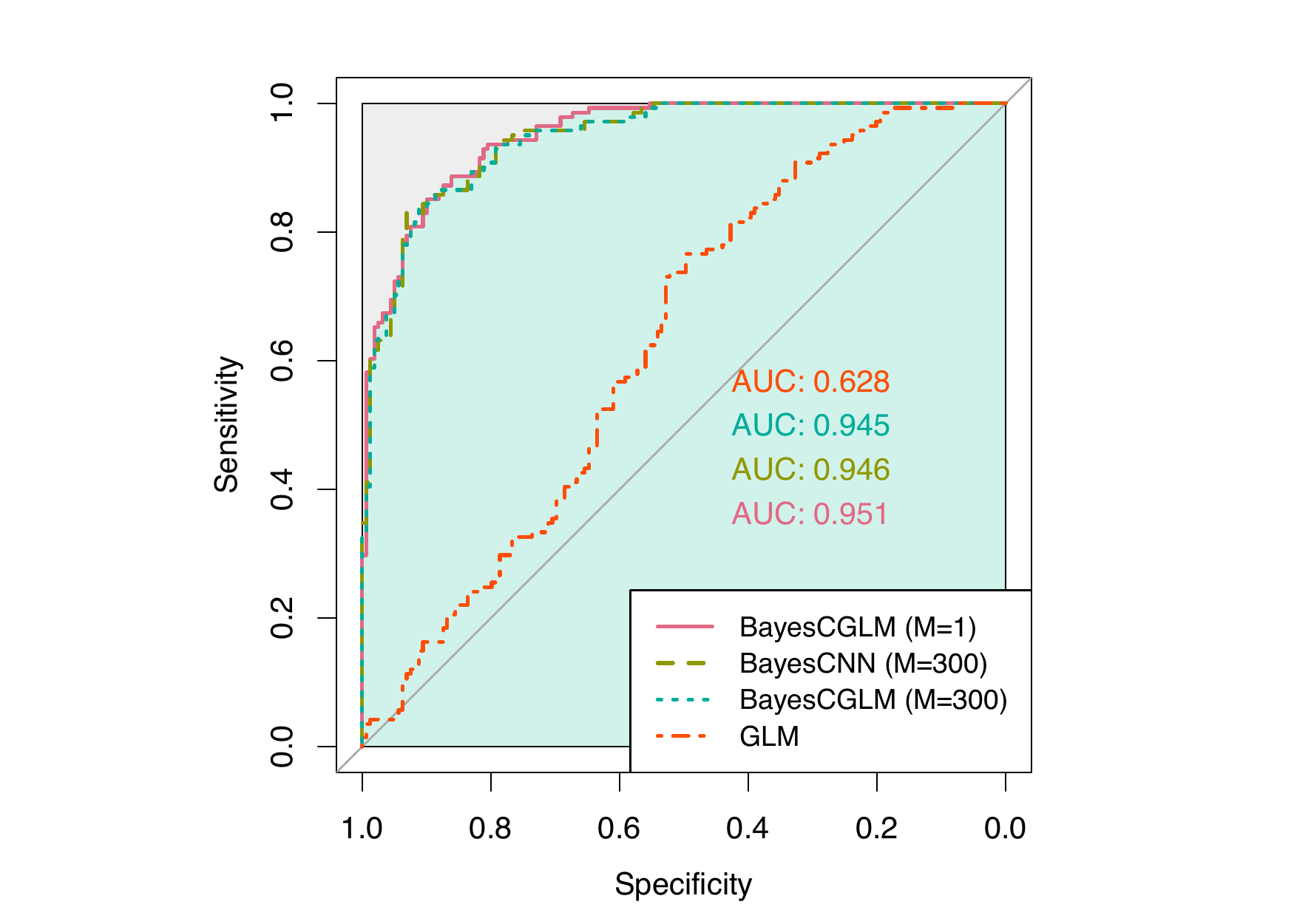}
  \caption{ROC plot}
  \label{simul_pred_binary2}
\end{subfigure}
\caption{(a) The estimated probability of $y=1$ (blue solid line) and the observed binary responses (black dots). (b) The ROC curve and AUC obtained by each method. BayesCGLM with $M=1$ outperforms the other methods.}
\label{simul_pred_binary}
\end{figure}

We now describe the simulation study results when $\mathbf{Y}$ is in the form of binary response, i.e., $\mathbf{Y} \sim \text{Bernoulli}(\bm{\lambda})$. Figure \ref{simul_pred_binary} compares the true response and the estimated probabilities (for the response to be 1), which show a good agreement. We observe that the predicted probability shows higher values for the binary responses with a true value of 1. The area under the curve (AUC) from the receiver operating characteristic curve (ROC) is about 0.85, which is reasonably close to 1 (Figure \ref{simul_pred_binary2}).

In Table~\ref{simul_binary}, we observe that parameter estimates from BayesCGLM are accurate, and the coverages are close to the 95\% nominal rate even with a single MC dropout ($M=1$). On the other hand, the estimates from BayesCNN and GLM are biased. Deep learning methods (BayesCGLM and BayesCNN) show much better prediction performance than GLM because they can extract features from image observations. 

\begin{table}[tt]
\centering
\caption[]{Inference results for the simulated binary datasets with the sample of size 500. For all methods, posterior mean, accuracy, recall, precision, and computing time (min) are averaged from 300 repeated simulations. Estimation coverages are also reported. The numbers in the parentheses indicate standard deviations obtained from the repeated simulations.}
\label{simul_binary}
\begin{tabular}{l l c c c c}
\hline
 \multicolumn{2}{c}{} & \multicolumn{2}{c}{\textbf{BayesCGLM}} & \textbf{BayesCNN} & \textbf{GLM} \\ 
\cmidrule(lr){3-4} \cmidrule(lr){5-5}
 & &  $M=300$ & $M= 1$ & $M=300$ \\
\hline
$\gamma_1$ & Mean & 1.029 (0.154) & 1.030  (0.154)  & 0.848 (0.117) & 0.389  (0.037) \\
$\gamma_2$ & Mean & 1.030 (0.151)  & 1.024 (0.155) & 0.851 (0.120) & 0.388 (0.081) \\
& Coverage & 0.954 & 0.952 & -  & 0 \\  
Prediction & Accuracy & 0.869 (0.019) & 0.857 (0.022)  & 0.850 (0.022) & 0.600 (0.028)   \\
& Recall & 0.869 (0.031) & 0.855 (0.033) & 0.851 (0.036) & 0.600(0.066) \\
& Precision & 0.868 (0.030) & 0.858 (0.033) & 0.863 (0.029) &  0.602 (0.043)  \\
Time (min) & & 37.028  &  17.661 & 15.022 &  0.003\\
\hline
\end{tabular}
\end{table}

We also conduct a simulation study with samples of size 5,000. We use the same CNN structure and hyperparameters as before. Table~\ref{simul_binary_bigsample} indicates that BayesCGLM and BayesCNN show comparable prediction performance, while GLM does not work well. We observe that BayesCGLM provides the most accurate regression coefficient estimates with reasonable uncertainty quantification.

\begin{table}[tt]
\centering
\caption{Inference results for the simulated binary datasets with the sample of size 5,000. For all methods, posterior mean, accuracy, recall, precision, and computing time (min) are averaged from 300 repeated simulations. Estimation coverages are also reported. The numbers in the parentheses indicate standard deviations obtained from the repeated simulations.}
\label{simul_binary_bigsample}
\begin{tabular}{l l  c c c c}
\hline
 \multicolumn{2}{c}{} & \multicolumn{2}{c}{\textbf{BayesCGLM}} & \textbf{BayesCNN} & \textbf{GLM} \\ 
\cmidrule(lr){3-4} \cmidrule(lr){5-5}
 & &  $M=300$ & $M= 1$ & $M=300$ \\
\hline
$\gamma_1$ & Mean & 0.968 (0.066) & 0.958  (0.080)  & 0.920 (0.06) & 0.368  (0.037) \\
& Coverage & 0.920 & 0.830  & - &  0\\  
$\gamma_2$ & Mean & 0.964 (0.065) & 0.954 (0.077) & 0.915 (0.06) & 0.382 (0.040)  \\
& Coverage & 0.917 & 0.837 &  - & 0 \\  
Prediction & Accuracy & 0.871 (0.009) & 0.865 (0.011)  & 0.870 (0.009) & 0.602 (0.012)   \\
& Recall & 0.871 (0.014) & 0.866 (0.017) & 0.870 (0.017) & 0.602 (0.030) \\
& Precision & 0.871 (0.013) & 0.865 (0.015) & 0.871 (0.016) &  0.602 (0.019)  \\
Time (min) & & 86.603 & 25.341 &  22.177 & 0.004 \\
\hline
\end{tabular}
\end{table}

\begin{figure}[htbp]
\begin{subfigure}{.5\textwidth}
  \centering
  \includegraphics[width=.99\linewidth]{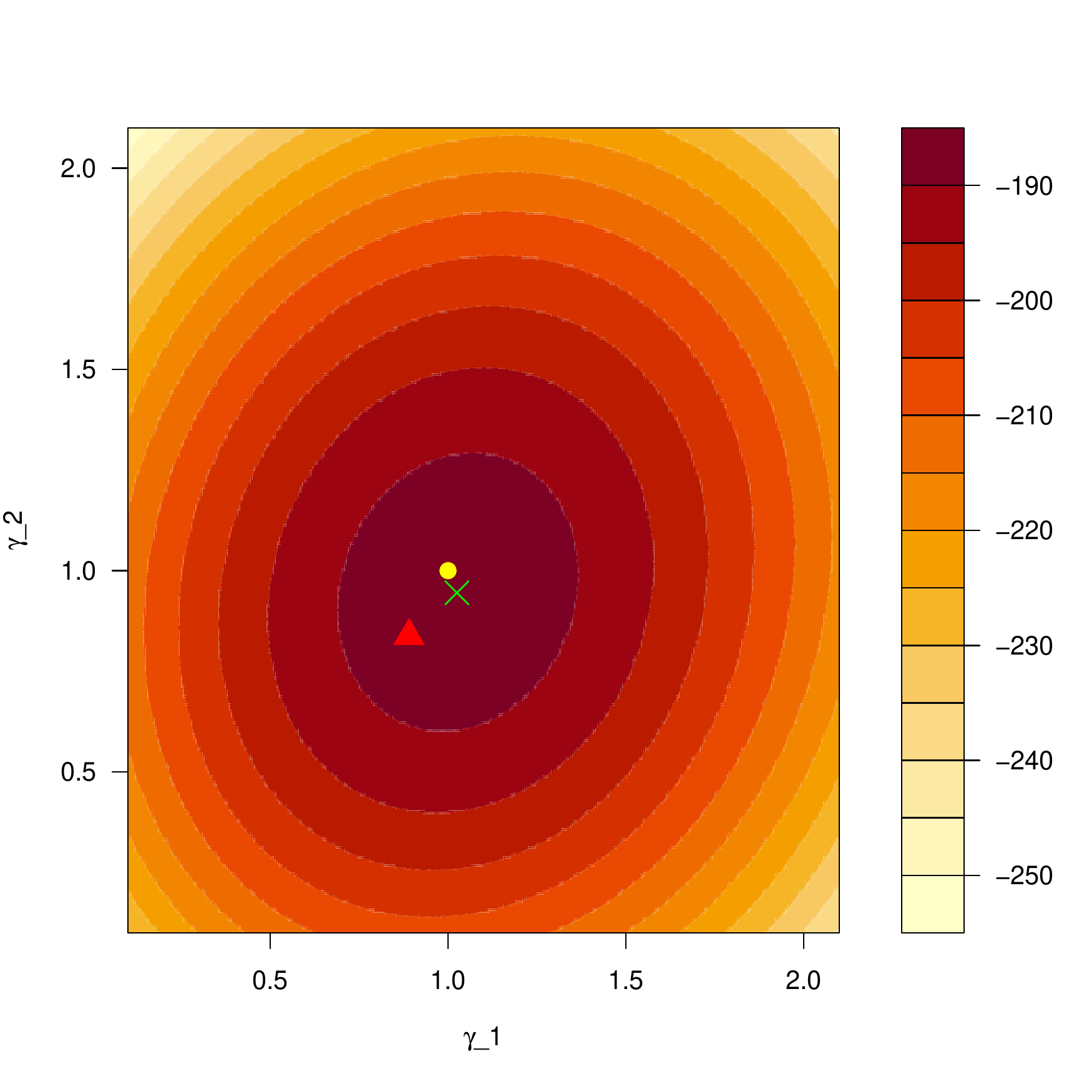}
  \caption{The profile log-likelihood in BayesCNN}
  \label{BayesCNN_loglike}
\end{subfigure}%
\begin{subfigure}{.5\textwidth}
  \centering
  \includegraphics[width=.99\linewidth]{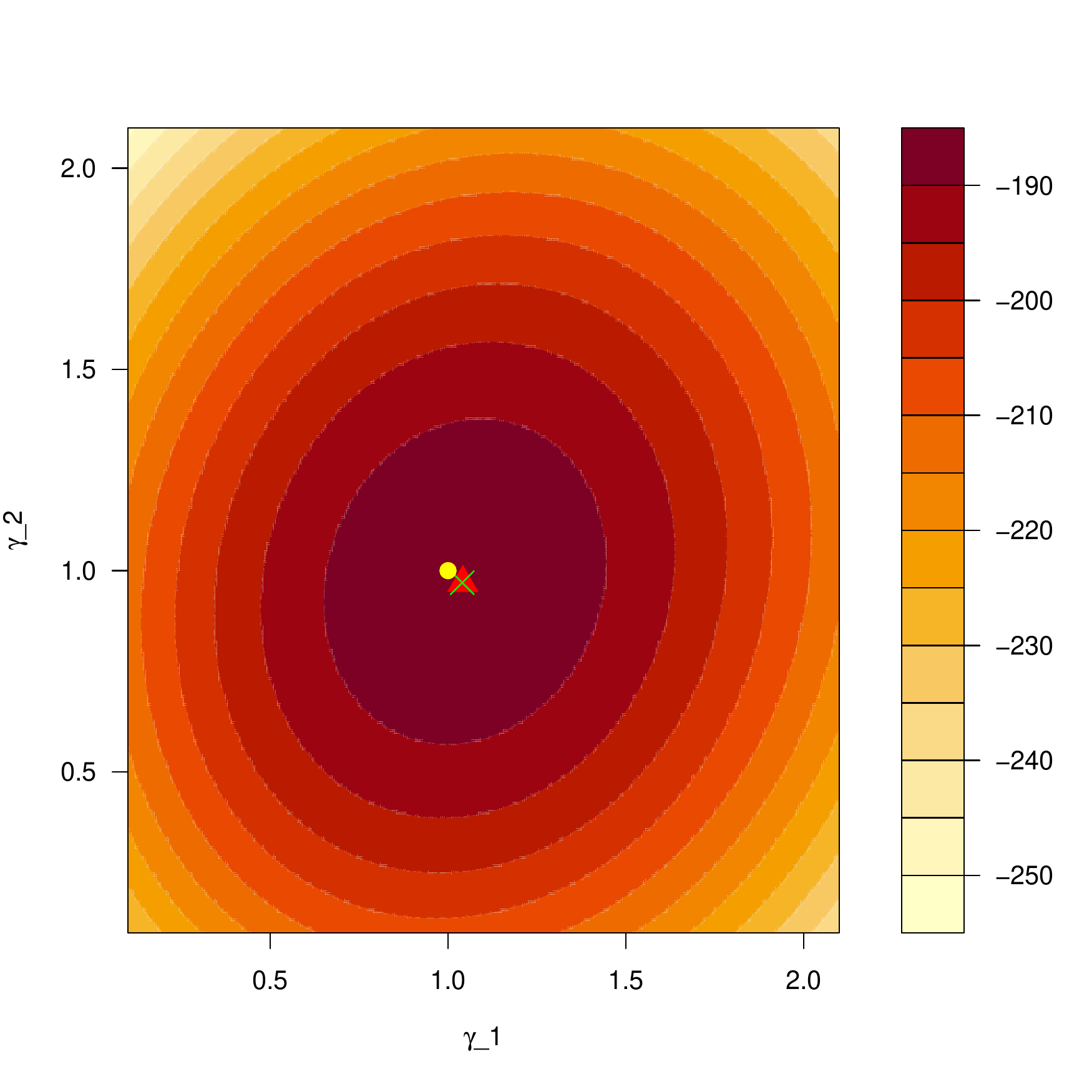}
  \caption{The profile log-likelihood in BayesCGLM}
  \label{BayesCGLM_loglike}
\end{subfigure}
\caption{The profile log-likelihood for $
\bm{\gamma}$ given other parameters fitted by CNN. The yellow circles represent the true coefficient $\bm{\gamma}=(1,1)$, the green crosses represent the profile likelihood estimates, and the red triangles represent the Bayes estimates obtained by BayesCNN and BayesCGLM, respectively. The estimate obtained by BayesCGLM is closer to the true $\bm{\gamma}$.}
\label{loglikelihood_contour}
\end{figure}

Figure~\ref{loglikelihood_contour} illustrates the profile log-likelihood surfaces over $\bm{\gamma}$ based on a single replicate from our simulation studies. As we described in the main manuscript, the SGD cannot guarantee the convergence of individual parameters, while our method can alleviate this issue.

\subsection{Poisson Case} \label{Poisson_case}
\begin{figure}[htbp]
\begin{center}
\includegraphics[width = 1\textwidth]{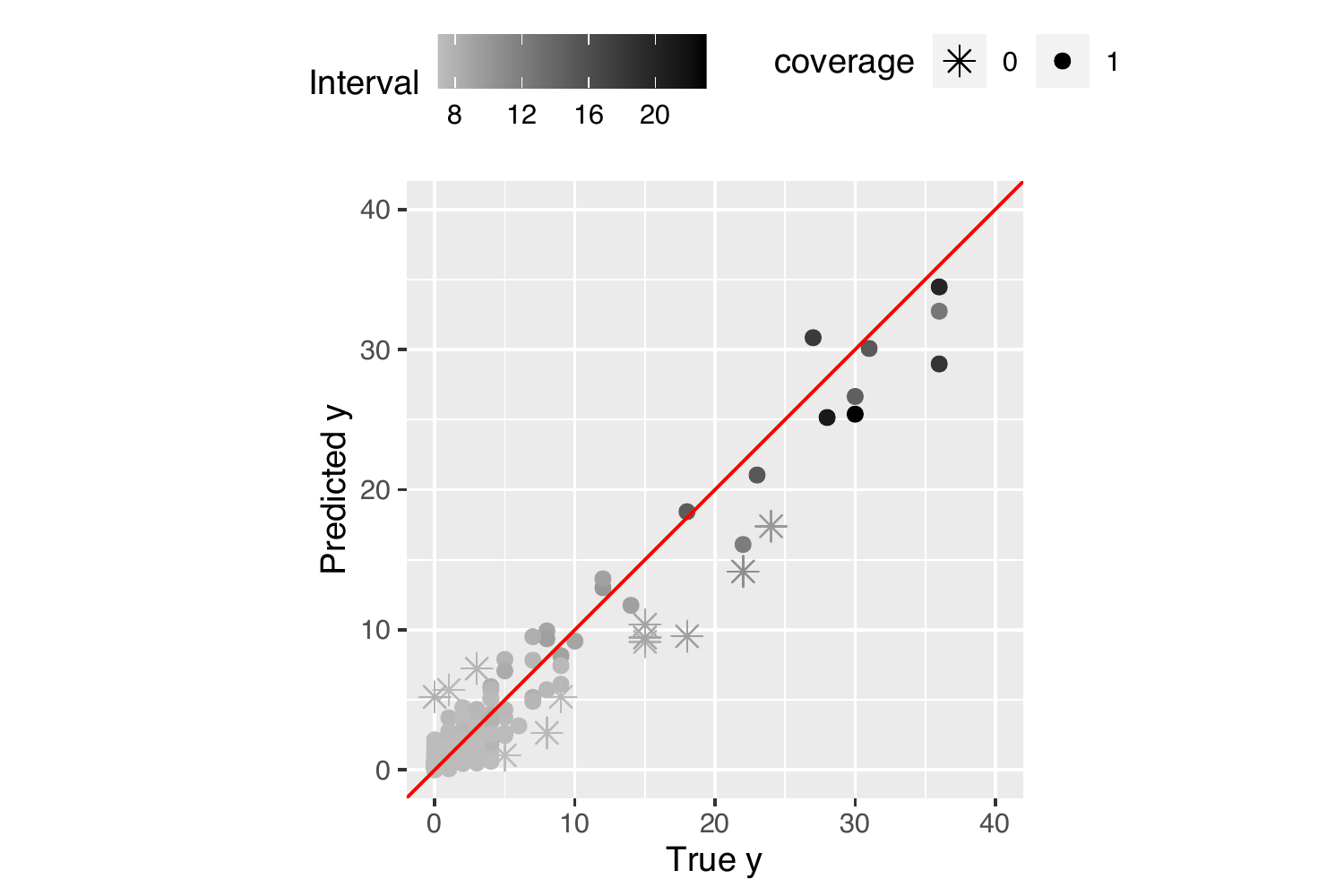}
\end{center}
\caption[]{Comparison between the true and predicted Poisson responses from BayesCGLM. Circles (asterisks) represent that the true responses are (not) covered in 95\% HPD intervals.}
\label{simul_pred_poisson}
\end{figure}

We describe the simulation study results when $\mathbf{Y}$ is generated from the Poisson distribution, i.e., $\mathbf{Y} \sim \text{Poisson}(\bm{\lambda})$.
Figure \ref{simul_pred_poisson} shows that the true and predicted responses are well aligned for a simulated dataset except for some extreme values. We observe that HPD prediction intervals include the true count response as well. 

Table~\ref{simul_poisson} indicates that BayesCGLM shows comparable (or better) prediction performance compared to BayesCNN while providing uncertainties in estimates. We observe that the coverage of the credible and prediction intervals becomes closer to the 95\% nominal rate with the repeated MC dropout sampling ($M=300$), compared to a single MC dropout sampling ($M=1$). Compared to BayesCNN, BayesCGLM with $M=300$ results in better point estimates for the parameters $\bm{\gamma}$ and point predictions for the response $\mathbf{Y}$. Compared to GLM, BayesCGLM with $M=300$ results in better empirical coverage for the credible intervals for $\bm{\gamma}$ as well as better prediction accuracy and empirical coverage for the prediction intervals for $\mathbf{Y}$.

\begin{table}[tt]
\centering
\caption[]{Inference results for the simulated Poisson datasets with the sample of size $500$. For all methods, posterior mean, RMSPE, and computing time (min) are averaged from 300 repeated simulations. Estimation and prediction coverages are also reported. The numbers in the parentheses indicate standard deviations obtained from the repeated
simulations.}
\label{simul_poisson}
\begin{tabular}{l l c c c c}
\hline
 \multicolumn{2}{c}{} & \multicolumn{2}{c}{\textbf{BayesCGLM}} & \textbf{BayesCNN} & \textbf{GLM} \\ 
\cmidrule(lr){3-4} \cmidrule(lr){5-5}
 & &  $M=300$ & $M= 1$ & $M=300$ \\
\hline
$\gamma_1$ & Mean & 0.987 (0.027) & 0.989 (0.031)  & 0.910 (0.106) &  0.997 (0.058)   \\
& Coverage & 0.936  & 0.862 & - & 0.542 \\  
$\gamma_2$ & Mean & 0.986 (0.026) & 0.989 (0.031)  & 0.911 (0.101) & 0.997 (0.059)  \\
& Coverage & 0.958 & 0.868 & - & 0.550 \\  
Prediction & RMSPE & 2.305 (1.610) & 2.687 (2.190) & 2.737 (2.537) & 4.364 (3.739) \\
& Coverage & 0.963 & 0.890  & 0.967 & 0.924 \\
Time (min) & & 22.730 &  15.285  & 13.753   & 0.004 \\ 
\hline
\end{tabular}
\end{table}

We compared the inferential results with a larger sample size (Table~\ref{simul_poisson_bigsample}). As in the previous examples, we conduct an additional study with samples of size 5,000. Both BayesCGLM and BayesCNN provide a comparable prediction performance in terms of RMSPE and coverage. Compared to the case with a smaller sample size, BayesCNN provides more accurate estimates of coefficients, indicating that the performance of the deep learning method improves with a larger sample size. BayesCGLM works well for both cases and can provide a reasonably accurate uncertainty quantification for regression coefficients.

\begin{table}[tt]
\centering
\caption{Inference results for the simulated Poisson datasets with the sample of size $5,000$. For all methods, posterior mean, RMSPE, and computing time (min) are averaged from 300 repeated simulations. Estimation and prediction coverages are also reported. The numbers in the parentheses indicate standard deviations obtained from the repeated simulations.}
\label{simul_poisson_bigsample}
\begin{tabular}{l l c c c c}
\hline
 \multicolumn{2}{c}{} & \multicolumn{2}{c}{\textbf{BayesCGLM}} & \textbf{BayesCNN} & \textbf{GLM} \\ 
\cmidrule(lr){3-4} \cmidrule(lr){5-5}
 & &  $M=300$ & $M= 1$ & $M=300$ \\
\hline
$\gamma_1$ & Mean &  0.999 (0.012) &  1.000 (0.012)  & 1.001 (0.012) & 1.001  (0.030) \\
& Coverage & 0.960 & 0.930 & - & 0.450  \\  
$\gamma_2$ & Mean & 0.999 (0.010) & 1.000 (0.011) & 1.001 (0.011) & 0.996 (0.032)  \\
& Coverage & 0.970 & 0.950 &  - & 0.370  \\  
Prediction & RMSPE & 1.927 (0.228) & 1.957 (0.261)  & 2.016 (0.288)  & 4.542 (1.555)  \\       
& Coverage & 0.967 & 0.943  & 0.944  & 0.968 \\    
Time (min) & & 117.742 & 66.607 & 63.864 &  0.005\\
\hline
\end{tabular}
\end{table}

\subsection{Comparison with Fully Bayesian Estimation}

In this section, we compare three different methods: (1) our two-stage approach (BayesDGLM), (2) Bayesian DNN (BayesDNN) \citep{gal2016dropout}, and (3) a fully Bayesian method (MCMC). Here, we refer to our method as BayesDGLM because standard NN is used to extract features instead of CNN for this example. Note that only the MCMC approach is exact without any approximations. We first generate input variables $\mathbf{X}=\lbrace \mathbf{x}_{n} \rbrace_{n=1}^{600} \in \mbR^{600 \times 3}$ and $\mathbf{Z}=\lbrace \mathbf{z}_{n} \rbrace_{n=1}^{600} \in \mbR^{600 \times 2}$, where individual elements are sampled from $N(0,1)$. Then the response variable $\mathbf{Y}=\lbrace y_{n} \rbrace_{n=1}^{600} \in \mbR^{600}$ is generated from $N(\mu_n,1)$. Here $\mu_n$ is obtained as 
\begin{equation}\begin{split}
    o_{n,1} &= \tanh(x_{n,1} w_1 + x_{n,2} w_2 + x_{n,3} w_3 + b_1), \\ 
    o_{n,2} &= \tanh(x_{n,1} w_4 + x_{n,2} w_5 + x_{n,3} w_6 + b_2), \\
    o_{n,3} &=  \tanh(x_{n,1} w_7 + x_{n,2} w_8 + x_{n,3} w_9 + b_3),\\
    \mu_n &= o_{n,1} w_{10} + o_{n,2} w_{11}+ o_{n,3} w_{12} + z_{n,1} \gamma_1 + z_{n,2} \gamma_2+  b_4,
\end{split}
\label{fullyBayes_mu}
\end{equation} 
where we use the weight parameters $(w_1,\cdots,w_{12}) =(1,2,1,2,3,1,0.5,0.8,1.1,0.5,1,1.5)$, bias parameters $(b_1,\cdots,b_4) = (0.2,0.2,0.2,0.3)$, and coefficients for the concatenated covariates as $(\gamma_1,\gamma_2)=(1,2)$.

For the approximated methods (BayesDGLM and BayesDNN), we use the same neural network architecture as in \eqref{fullyBayes_mu}. To train both methods, we use the Adam optimizer (learning rate $10^{-3}$) with a mean squared loss function, dropout rate 0.2, dropout sample $M=300$, batch size 10, and epoch size 10. For the fully Bayesian method, we use the standard normal priors for all parameters. The MCMC algorithm was run for 10,000 iterations, with the first 5,000 iterations discarded for burn-in. For all three methods, we use the first $N=500$ observations for training and the remaining $N_{\text{test}}=100$ for testing.

\begin{table}[ht]
\centering
\caption{Inference results for the simple neural network example. For all methods, posterior mean, RMSPE, and computing time (sec) are averaged from 100 repeated simulations. Estimation and prediction coverages are also reported. The numbers in the parentheses indicate standard deviations obtained from the repeated simulations.}
\label{fully_bayes_gaussian}
\begin{tabular}{l l c c c}
& & \textbf{BayesDGLM}&\textbf{BayesDNN}& \textbf{MCMC}  \\ & &  $M=300$ & $M=300$ & \\
\hline
$\gamma_1$ & Mean & 1.005 (0.042) &  1.005 (0.041)   & 0.994 (0.043)  \\
& Coverage & 0.960 & - & 0.960  \\  
$\gamma_2$ & Mean & 1.992 (0.047) & 1.994 (0.048) & 1.994 (0.046)  \\
& Coverage & 0.920 & - & 0.920 \\  
Prediction & RMSPE & 1.115 (0.077)  & 1.129 (0.080)  & 1.104 (0.077)  \\       
& Coverage & 0.925 & 0.922  & 0.937  \\    
Time (sec) & & 27.311 & 9.642  & 128.520 \\
\hline
\end{tabular}
\end{table}

Table~\ref{fully_bayes_gaussian} indicates that the fully Bayesian method provides the smallest RMSPE with coverage close to 95\% nominal rate, though it is computationally more expensive. Both approximated methods provide similar prediction results to the fully Bayesian method. However, BayesDNN \citep{gal2016dropout} cannot quantify uncertainties in parameter estimates, while our BayesDGLM provides similar coverages compared to the MCMC algorithm. In summary, under this simple NN structure, BayesDGLM provides comparable inference results to the exact method within a faster computing time. Note that for complex DNNs (or CNNs), implementing the MCMC algorithm is infeasible; in such a case, our method would be a practical option.

\section{Malaria Cases in the African Great Lakes Region}
 Malaria is a parasitic disease that can lead to severe illnesses and even death. The burden of malaria is disproportionately high in the African regions according to the World Health Organization \citep[WHO;][]{world2022world}. Therefore, studying malaria incidence in this region is of significant public health interest for effective control interventions. Previously, \citet{gopal2019characterizing} studied the malaria incidence rate in Kenya with several environmental variables. However, their work focuses on exploratory data analysis rather than constructing spatial models with adequate uncertainty quantification. On the other hand, we incorporate spatial correlation into the model through the extracted feature and quantify uncertainties in parameter estimation and prediction. Furthermore, we expand the analysis to the entire African Great Lakes region by compiling malaria incidence data from the Demographic and Health Surveys of 2015 \citep{icf2015demographic}. The dataset contains malaria incidence (counts) from 4,741 GPS clusters in nine contiguous countries in the African Great Lakes region: Burundi, the Democratic Republic of the Congo, Malawi, Mozambique, Rwanda, Tanzania, Uganda, Zambia, and Zimbabwe. We use the average annual rainfall, the vegetation index of the region, and the proximity to water as spatial covariates. We use $N = 3,500$ observations to fit the model and save $N_{\text{test}} = 1,241$ observations for validation.

Here, we use a spatial basis function matrix as the correlated input $\mathbf{X}$. We place $25\times 25$ knots over the rectangular domain that covers the irregularly shaped spatial domain. Then we chose 239 knots, which are located inside the African Great Lakes region. We use the thin plate splines basis defined as $\mathbf{X}_{j}(\mathbf{s}) = ||\mathbf{s}-\mathbf{u}_{j}||^{2}\log(||\mathbf{s}-\mathbf{u}_{j}||)$, where $\mathbf{s}$ is a spatial location and $\lbrace \mathbf{u}_{j} \rbrace_{j=1}^{239}$ is a set of knots placed over our region of interest. From this, we can construct basis functions for each observation, resulting in a $\mathbf{X} \in \mbR^{3,500 \times 239}$ input design matrix. Since each observation is associated with a $239$-dimensional basis vector, we apply 1-D convolution to extract features. We first train BayesCNN with inputs $\mathbf{X}, \mathbf{Z}$ and the count response $\mathbf{Y}$. From this, we extract a feature design matrix $\bm{\Phi} \in \mbR^{3,500 \times 16}$ from the last layer of the fitted BayesCNN. With covariate matrix $\mathbf{Z} \in \mbR^{3,500 \times 3}$, we fit Poisson GLMs by regressing $\mathbf{Y}$ on $[\mathbf{Z}, \bm{\Phi}^{(m)}] \in \mbR^{3,500 \times 19}$ for each $m$; here, we have used $M=500$. 

We compare our method with a spatial basis regression model \citep[cf.][]{lee2020scalable} as
\[
g(E[\mathbf{Y}|\mathbf{Z},\mathbf{X}]) = \mathbf{Z}\bm{\gamma} + \mathbf{X}\bm{\delta}, 
\]
where $\mathbf{X} \in \mbR^{3,500 \times 239}$ is the same thin plate basis matrix that we used in BayesCGLM. To fit a hierarchical spatial regression model, we use {\tt{nimble}} in {\tt{R}} (https://cran.r-project.org/web/
packages/nimble). To guarantee convergence, an MCMC algorithm is run for 1,000,000 iterations, with 500,000 discarded for burn-in. Furthermore, we also implement BayesCNN as in the simulated examples. Table~\ref{malaria_table} indicates that vegetation, water, and rainfall variables have positive relationships with malaria incidence in deep learning methods. All coefficients in our model are statistically significant based on the highest posterior density (HPD) intervals. Previous studies have also shown that these covariates affect malaria incidence in Uganda and Kenya \citep{gopal2019characterizing, okiring2021associations}. However, we expand our study scope to include the African Great Lakes region, which to our knowledge, has not been studied before. Furthermore, our model can capture spatial correlation, while previous studies used basic regression models. BayesCGLM shows comparable prediction performance compared to BayesCNN. Although BayesCNN can provide prediction uncertainties, it cannot quantify uncertainties of the estimates. Although we can obtain HPD intervals from spatial regression, it provides a higher RMSPE and lower coverage than other deep learning methods. 

\begin{table}[tt]
\centering
\caption[]{Inference results for the malaria dataset from different methods. For all methods, the posterior mean of $\bm{\gamma}$, 95\% HPD interval, RMSPE, prediction coverage, and computing time (min) are reported in the table.}
\label{malaria_table}
\begin{tabular}{l l c c c}
& & \textbf{BayesCGLM} &  \textbf{BayesCNN} &\textbf{Spatial model} \\ 
& & $M=500$ & $M=500$ & \\
\hline
$\gamma_1$ (vegetation index) & Mean  & 0.099  & 0.103 &  0.115  \\
       & 95\% Interval  & $(0.092, 0.107)$  & - &  $(0.111, 0.118)$  \\
$\gamma_2$ (proximity to water) & Mean  & 0.074   & 0.058  & $-0.269$ \\
   & 95\% Interval  & $(0.068, 0.080)$  & - & $(-0.272, -0.266)$ \\
$\gamma_3$ (rainfall) & Mean & 0.036  & 0.027 & $-0.122$ \\
     & 95\% Interval & $(0.027, 0.045)$& -  &$(-0.126, -0.117)$ \\
Prediction &  RMSPE &  27.438 &  28.462 & 42.393  \\ 
& Coverage & 0.950 & 0.947 & 0.545 \\
Time (min) & & 57.518 & 30.580 & 41.285 \\
\hline 
\end{tabular}
\end{table}

\begin{figure}[htbp]
\begin{center}
\includegraphics[width = 0.7\textwidth,trim={1.5mm 1.5mm 1.5mm 1.5mm}]{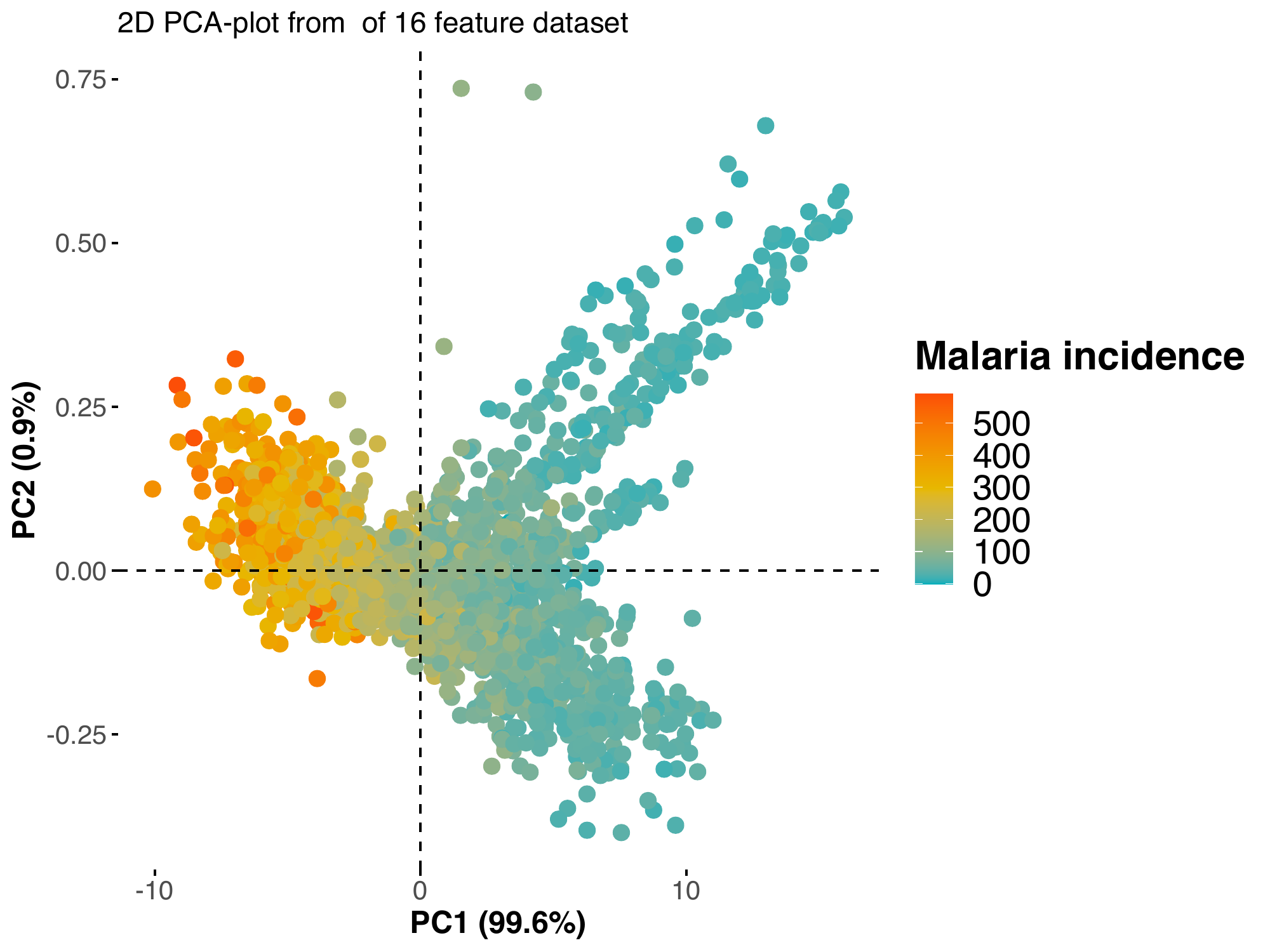}
\end{center}
\caption[]{Score plots for malaria incidence dataset.}
\label{Phi_malaria}
\end{figure}

Figure~\ref{Phi_malaria} shows a score plot with the first and second principal components of $\bm\Phi$. We observe a negative correlation between malaria incidence and the first principal component (99.6\% explained variability). This implies that the extracted feature is an informative summary statistic for predicting malaria incidence. Incorporating the feature information into the model, BayesCGLM can provide the most accurate prediction results with adequate uncertainty quantification.

\begin{figure}[htbp]
\begin{center}
\includegraphics[width = 0.9\textwidth]{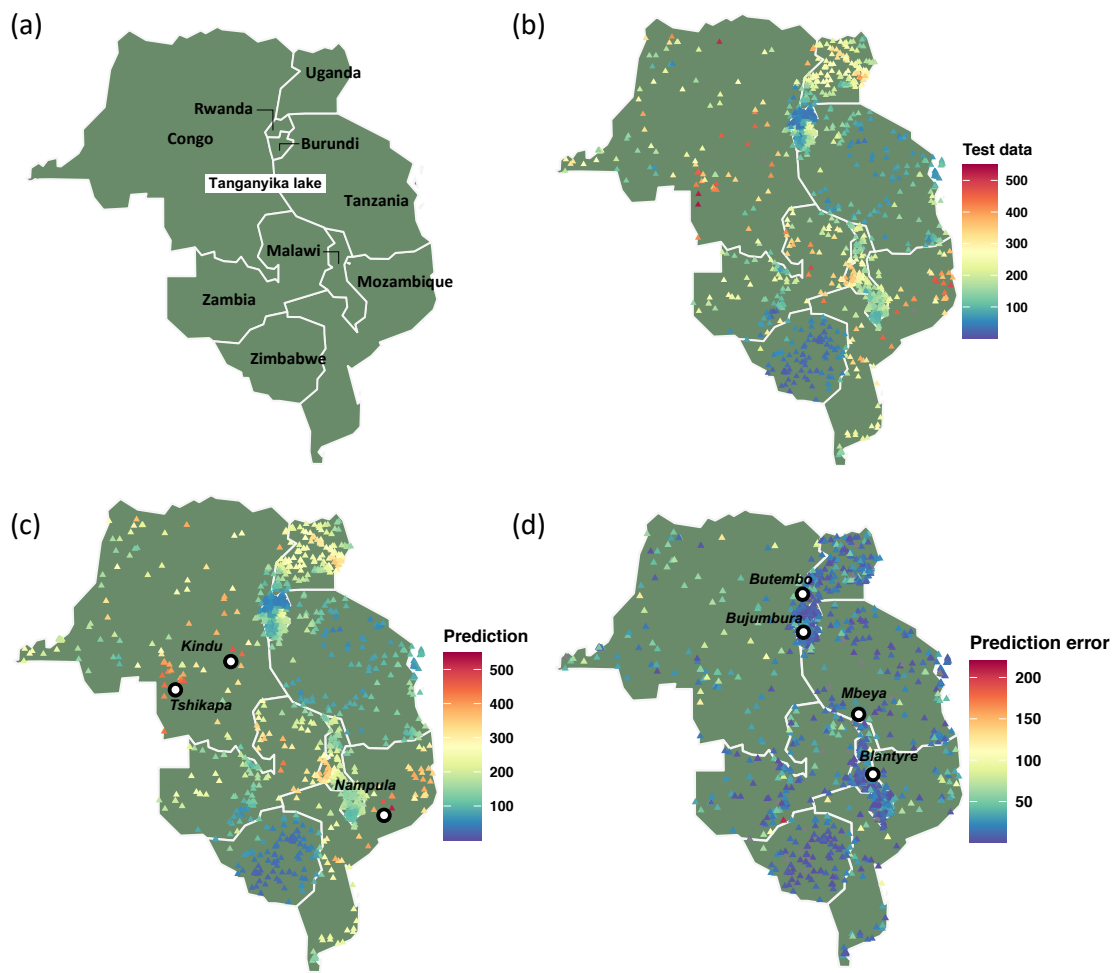}
\end{center}
\caption[]{(a) The nine countries in the African Great Lakes region. (b) The true malaria incidence from the test dataset. (c) Prediction results of malaria incidence. Circles indicate the cities with high malaria incidence (hotspots). (d) Prediction error of malaria incidence. Circles indicate the cities with low prediction error.
}
\label{fig1_malaria}
\end{figure}

Figure~\ref{fig1_malaria} shows that the true and predicted incidence from BayesCGLM have similar spatial patterns. Specifically, BayesCGLM can detect several hotspots well, including Tshikapa and Kindu in Congo, and Nampula in Mozambique. In Figure~\ref{fig1_malaria}, we observe that prediction errors are low in regions near Lake Tanganyika, such as Butembo in Congo, Bujumbura in Burundi, Mbeya in Tanzania, and Blantyre in Malawi. On the other hand, prediction errors become larger in areas further away from Lake Tanganyika. This is mainly due to the limited number of observed cases in those regions; the sparse observation in regions further away from Lake Tanganyika leads to higher prediction errors. Therefore, caution is necessary when policymakers predict incidence in such regions.

\newpage
\section{CNNStructures}\label{Sec:CNNstruc}
Convolution layer, max pooling, flatten, dense, and concatenate are the building blocks of CNNs. Convolution layers apply a set of filters to an input image to generate a set of feature maps that capture spatially local patterns. Max pooling layers downsample the feature maps by extracting the maximum value within each local region, reducing the spatial dimension of the output and enhancing translation invariance. Flatten layers convert the multidimensional feature maps into a one-dimensional vector, which is fed into fully connected dense layers that perform non-linear transformations of the input variables. Concatenate layers merge multiple inputs along a specified axis, combining different sets of input variables. Here, we use concatenate layers to combine image features and covariates.
\subsection{Simulations}\label{Sec:CNNstrucSimu}

\paragraph{Gaussian Data}
\begin{itemize}
\item Optimizer: Adam optimizer
\item Learning rate: $1e-4$
\item Loss function: Mean squared error loss $\frac{1}{N}\sum_{i=1}^{N}(y_i - \widehat{y_i})^2$
\item Batch size: 32
\item Epoch size: 300
\end{itemize}

\begin{table}[tt]
\centering
\caption{Summary of 2D-CNN configurations for Gaussian simulation}
\begin{tabular}{c c c c c c}
\textbf{Layer type} &\textbf{Dimension}& \textbf{Kernel size} & \textbf{Strides} & \textbf{Activation function}  \\ \Hline
Convolution layer& $8\times 8$ & $4\times4$ & $2\times2$ & ReLU \\
Dropout ($\psi=0.2$) & - & - & - & - \\  
Max pooling & $2\times 2$ & - & - & -  \\
\hline
Convolution layer & $16\times 16$  & $3\times 3$ & $2\times2$ & Softmax \\   
Dropout ($\psi=0.2$) & - & - & - & - \\       
Max pooling & $2\times 2$ & - & -  & -\\    
\hline
Flatten &  - & -  & - & - \\    
Dense & $1\times 32$  & - & - & ReLU \\  
Dropout ($\psi=0.2$) & - & - & - & - \\  
Dense & $1\times 16$  & - & - & ReLu \\
Dropout ($\psi=0.2$) & - & - & - & - \\
Dense & $1\times 16$  & - & - & Softplus \\ 
Dropout ($\psi=0.2$) & - & - & - & - \\ 
\hline
Concatenate & $\mathbf{Z}$ \\ 
Dense & $1\times 1$  & - & - & Linear \\ 
\hline
\end{tabular}
\end{table}

\paragraph{Binary Data}
\begin{itemize}
\item Optimizer: Adam optimizer
\item Learning rate: $1e-4$
\item Loss function: Binary cross entropy loss $- \frac{1}{N} \sum_{i=1}^{N} y_i \log(\widehat{y_i})+ (1-y_i)\log(1-\widehat{y_i})$
\item Batch size: 3
\item Epoch size: 2,000
\end{itemize}

\begin{table}[tt]
\centering
\caption{Summary of 2D-CNN configurations for binary simulation}
\begin{tabular}{c c c c c c}
\textbf{Layer type} &\textbf{Dimension}& \textbf{Kernel size} & \textbf{Strides} & \textbf{Activation function}  \\ \Hline
Convolution layer& $16\times 16$ & $3\times3$ & $1\times 1$ & Softmax \\
Dropout ($\psi=0.25$) & - & - & - & - \\  
Max pooling & $2\times 2$ & - & - & -  \\
\hline
Convolution layer & $32\times 32$  & $3\times 3$ & $1\times 1$ & Softmax \\   
Dropout ($\psi=0.25$) & - & - & - & - \\       
Max pooling & $2\times 2$ & - & -  & -\\    
\hline
Flatten &  - & -  & - & - \\    
Dense & $1\times 16$  & - & - & ReLU \\  
Dropout ($\psi=0.25$) & - & - & - & - \\  
Dense & $1\times 8$  & - & - & Linear \\
Dropout ($\psi=0.25$) & - & - & - & - \\
\hline
Concatenate & $\mathbf{Z}$ \\ 
Dense & $1\times 1$  & - & - & Sigmoid \\ 
\hline
\end{tabular}
\end{table}

\paragraph{Poisson Data}
\begin{itemize}
\item Optimizer: Adam optimizer
\item Learning rate: $1e-3$
\item Loss function: Poisson loss $\frac{1}{N}\sum_{i=1}^{N}(\widehat{y_i}-y_i \log (\widehat{y_i}+\epsilon))$; here $\epsilon >0$ is added to avoid numerical instability when $\widehat{y}=0$.
\item Batch size: 3
\item Epoch size: 2,000
\end{itemize}

  \begin{table}[tt]
\centering
\caption{Summary of 2D-CNN configurations for Poisson simulation}
\begin{tabular}{c c c c c c}
\textbf{Layer type} &\textbf{Dimension}& \textbf{Kernel size} & \textbf{Strides} & \textbf{Activation function}  \\ \Hline
Convolution layer& $8\times 8$ & $4\times 4$ & $2\times 2$ & Softmax \\
Dropout ($\psi=0.2$) & - & - & - & - \\  
Max Pooling & $2\times 2$ & - & - & -  \\
\hline
Convolution layer & $32\times 32$  & $3\times 3$ & $1\times 1$ & Softmax \\   
Dropout ($\psi=0.2$) & - & - & - & - \\       
Max pooling & $2\times 2$ & - & -  & -\\    
\hline
Flatten &  - & -  & - & - \\    
Dense & $1\times 32$  & - & - & Softplus \\  
Dropout ($\psi=0.2$) & - & - & - & - \\  
Dense & $1\times 16$  & - & - & Linear \\
Dropout ($\psi=0.2$) & - & - & - & - \\
\hline
Concatenate & $\mathbf{Z}$ \\ 
Dense & $1\times 1$  & - & - & Exponential \\ 
\hline
\end{tabular}
\end{table}

\subsection{Real Data Applications}\label{Sec:CNNstrucReal}

\paragraph{Malaria Incidence}
\begin{itemize}
\item Optimizer: Adam optimizer
\item Learning rate: $1e-4$
\item Loss function: Poisson loss $\frac{1}{N}\sum_{i=1}^{N}(\widehat{y_i}-y_i \log (\widehat{y_i}+\epsilon))$; here $\epsilon >0$ is added to avoid numerical instability when $\widehat{y}=0$.
\item Batch size: 10
\item Epoch size: 2500
\end{itemize}

  \begin{table}[tt]
\centering
\caption{Summary of 1D-CNN configurations for malarial incidence}
\begin{tabular}{c c c c c c}
\textbf{Layer type} &\textbf{Dimension}& \textbf{Kernel size} & \textbf{Strides} & \textbf{Activation function}  \\ \Hline
Convolution layer& $1\times 32$ & $3$ & $1$ & TanH \\
Dropout($\psi=0.25$) & - & - & - & - \\  
Max pooling & $1\times 1$ & - & - & -  \\
\hline
Convolution layer & $1\times 64$  & $3$ & $1$ & TanH \\   
Dropout ($\psi=0.25$) & - & - & - & - \\       
Max pooling & $2\times 2$ & - & -  & -\\    
\hline
Flatten &  - & -  & - & - \\    
Dense & $1\times 32$  & - & - & ReLU \\  
Dropout ($\psi=0.25$) & - & - & - & - \\  
Dense & $1\times 16$  & - & - & Linear \\
Dropout ($\psi=0.25$) & - & - & - & - \\
\hline
Concatenate & $\mathbf{Z}$ \\ 
Dense & $1\times 1$  & - & - & Exponential \\ 
\hline
\end{tabular}
\end{table}

\paragraph{Brain Tumor MRI Images}
\begin{itemize}
\item Optimizer: Adam optimizer
\item Learning rate: $1e-4$
\item Loss function: Binary cross entropy loss $- \frac{1}{N} \sum_{i=1}^{N} y_i \log(\widehat{y_i})+ (1-y_i)\log(1-\widehat{y_i})$
\item Batch size: 3
\item Epoch size: 5
\end{itemize}

  \begin{table}[tt]
\centering
\caption{Summary of 2D-CNN configurations for brain tumor MRI images}
\begin{tabular}{c c c c c c}
\textbf{Layer type} &\textbf{Dimension}& \textbf{Kernel size} & \textbf{Strides} & \textbf{Activation function}  \\ \Hline
Convolution layer& $64\times 64$ & $3 \times 3 $ & $1 \times 1$ & ReLU \\
Dropout ($\psi=0.25$) & - & - & - & - \\  
Max pooling & $2\times 2$ & - & - & -  \\
\hline
Convolution layer & $32\times 32$  & $3 \times 3$ & $1\times 1$ & ReLU \\   
Dropout ($\psi=0.25$) & - & - & - & - \\       
Max pooling & $2\times 2$ & - & -  & -\\    
\hline
Flatten &  - & -  & - & - \\    
Dense & $1\times 16$  & - & - & Linear \\  
Dropout ($\psi=0.25$) & - & - & - & - \\  
\hline
Concatenate & $\mathbf{Z}$ \\ 
Dense & $1\times 1$  & - & - & Sigmoid \\ 
\hline
\end{tabular}
\end{table}

\paragraph{fMRI Data for Anxiety Score}
\begin{itemize}
\item Optimizer: Adam optimizer
\item Learning rate: $1e-3$
\item Loss function: Mean squared error loss $\frac{1}{N}\sum_{i=1}^{N}(y_i - \widehat{y_i})^2$
\item Batch size: 3
\item Epoch size: 500
\end{itemize}

  \begin{table}[tt]
\centering
\caption{Summary of 2D-CNN configurations for fMRI data}
\begin{tabular}{c c c c c c}
\textbf{Layer type} &\textbf{Dimension}& \textbf{Kernel size} & \textbf{Strides} & \textbf{Activation function}  \\ \Hline
Convolution layer& $8\times 8$ & $3 \times 3 $ & $2 \times 2$ & ReLU \\
Dropout ($\psi=0.25$) & - & - & - & - \\  
Max pooling & $2\times 2$ & - & - & -  \\
\hline
Convolution layer & $16\times 16$  & $3 \times 3$ & $2\times 2$ & ReLU \\   
Dropout ($\psi=0.25$) & - & - & - & - \\       
Max pooling & $2\times 2$ & - & -  & -\\    
\hline
Flatten &  - & -  & - & - \\    
Dense & $1\times 16$  & - & - & ReLU \\  
Dropout ($\psi=0.25$) & - & - & - & - \\  
Dense & $1\times 8$  & - & - & Softplus \\  
Dropout ($\psi=0.25$) & - & - & - & - \\  
\hline
Concatenate & $\mathbf{Z}$ \\ 
Dense & $1\times 1$  & - & - & Linear \\ 
\hline
\end{tabular}
\end{table}

\bibliography{Reference}